\documentclass[preprint,12pt,review]{elsarticle}

\usepackage{amssymb}
\usepackage{amsmath}
\usepackage{bm}
\usepackage{graphicx}
\usepackage{subcaption}
\usepackage{float}
\usepackage{xcolor}
\usepackage{hyperref}
\usepackage[margin=2cm]{geometry}
\usepackage{multirow}
\usepackage{siunitx}
\journal{International Journal of Fatigue}

\begin{document}
\begin{frontmatter}

\title{A computational framework for predicting the effect of surface roughness in fatigue}

\author{Sara Jim\'{e}nez-Alfaro\fnref{Imperial,OXFORD}}
\author{Emilio Mart\'{\i}nez-Pa\~neda\corref{cor1}\fnref{OXFORD}}
\ead{emilio.martinez-paneda@eng.ox.ac.uk}

\address[Imperial]{Department of Civil and Environmental Engineering, Imperial College London, London SW7 2AZ, UK}

\address[OXFORD]{Department of Engineering Science, University of Oxford, Oxford OX1 3PJ, UK}

\begin{abstract}
Surface roughness is a critical factor influencing the fatigue life of structural components. Its effect is commonly quantified using a correction coefficient known as the surface factor. In this paper, a phase field based numerical framework is proposed to estimate the surface factor while accounting for the stochastic nature of surface roughness. The model is validated against existing experimental data. Furthermore, we investigate the influence of key parameters on the fatigue life of rough surfaces, such as surface topology and failure strength. An important effect of surface roughness is observed when the average surface roughness increases and the correlation length of the surface profile decreases. This effect becomes more pronounced with higher failure strengths. 
\end{abstract}




\end{frontmatter}

\section*{Nomenclature}

\begin{table}[H]
\begin{tabular}{ll}
$a,b$ & Coefficients in Basquin's law \\
$\text{ACF}$ & Autocorrelation function \\
$\mathbf{b}$ & External volume force \\
$d_x$ & Length of a 1D profile sample where ACF is evaluated \\
$E$ & Young's modulus \\
$f$ & Fatigue degradation function \\
$g$ & Degradation function \\
$G_\text{c}$ & Critical energy release rate \\
$\mathcal{H}$ & History variable \\
$K_\text{IC}$ & Fracture toughness \\
\end{tabular}
\label{tab:nomenclature}
\end{table}

\begin{table}[H]
\begin{tabular}{ll}
$K_s$ & Surface factor \\
$\ell$ & Phase field length scale \\
$\ell_\text{cor}$ & Correlation length \\
$\ell_\text{Irwin}$ & Irwin's characteristic length \\
$\ell_\text{mesh}$ & Characteristic finite element length \\
$M$ & Number of simulations in the statistical sample \\
$\mathrm{ME}$ & Relative margin of error \\
$\mathbf{n}$ & outward normal vector \\
$N$ & Number of points on the real surface \\
$N_f$ & Mean of the statistical distribution of cycles to failure \\
$\mathbf{N_f}$ & Number of cycles to failure in the simulations of the statistical sample \\
$N_\square, \sigma_\square$ & Point in Basquin's law \\
$R$ & Stress ratio \\
$\mathbf{R}$ & Autocorrelation matrix \\
$R_\text{a}$ & Average surface roughness \\
$R_\text{q}$ & Root mean square roughness \\
$s$ & Standard deviation \\
$t$ & Time \\
$\mathbf{t}$ & External tractions per unit area \\
$\mathbf{u}$ & Displacement vector \\
$\mathbf{u}_\mathrm{e}$ &  Displacement field when the stress amplitude equals the endurance limit \\
$\mathbf{\bar{u}}$ & External applied displacement \\
$\mathbf{\dot{u}}$ & Velocity vector \\
$w$ & Dissipation function \\
$\mathbf{w}$ & Gaussian white noise \\
$\mathbf{x}_0$ & Initial position of the nodes in the mesh of the polished surface \\
$\mathbf{z}$ & Deviation in the normal direction of a set of points in a real surface from its nominal form \\
$\alpha$ & Fatigue history variable \\
$\alpha_\text{e}$ & Endurance threshold \\
$\alpha_\text{max}$ & Maximum value of fatigue history variable in the loading cycle \\
$\bar{\alpha}$ & Cumulated fatigue history variable \\
$\bar{\alpha}_\text{T}$ & Material parameter in the fatigue degradation function \\
\end{tabular}
\label{tab:nomenclature}
\end{table}

\begin{table}[H]
\begin{tabular}{ll}
$\Gamma$ & Crack surface \\
$\bm{\varepsilon}$ & Strain tensor \\
$\varepsilon_\text{I}, \varepsilon_\text{II}, \varepsilon_\text{III}$ & Principal strains \\
$\lambda, \mu$ & Lamé coefficients \\
$\nu$ & Poisson's ratio \\
$\sigma_\text{c}$ & Fracture (critical) strength \\
$\sigma_a$ & Stress amplitude \\
$\sigma_\text{e}$ & Endurance limit \\
$\sigma_\text{e}^\text{p}$ & Endurance limit of the polished specimen \\
$\sigma_\text{e}^\text{r}$ & Endurance limit of the rough specimen \\
$\sigma_\text{ULT}$ & Ultimate strength \\
$\phi$ & Phase field variable \\
$\psi$ & Strain energy density \\
$\psi_\text{e}$ & Elastic strain energy density \\
$\psi_\text{e0}^+$ & Tensile (undamaged) strain energy density \\
$\psi_\text{e0}^-$ & Compressive (undamaged) strain energy density \\
$\psi_\text{f}$ & Fracture energy density \\
$\Omega$ & Domain of the solid \\
$\partial \Omega $ & Contour of the solid \\
$\partial_t \Omega$ & Contour of the solid where external tractions per unit area are applied \\
$\partial_u \Omega$ & Contour of the solid where external displacements are applied \\
\end{tabular}
\label{tab:nomenclature}
\end{table}

\section{Introduction}\label{sec1}

Fatigue failure is recognised as the predominant cause of failure in load-bearing components \cite{Shigleys2011}. This phenomenon is influenced by multiple factors, including material composition, manufacturing processes, and environmental conditions. Among these, surface finish is particularly critical, since fatigue cracks often originate at the surface, where irregularities act as stress concentrators and promote crack initiation \cite{Suresh1998}. To address this influence, a correction factor known as the surface factor ($K_\mathrm{s}$) is commonly applied to degrade the fatigue properties obtained from polished specimens. First defined by Marin \cite{Marin1962}, the surface factor is expressed as the ratio of the fatigue strength at $10^6$ cycles of a rough specimen to that of a polished specimen. In the literature, a variety of models have been proposed to define the surface factor as a function of various properties, such as surface finish metrics, the type of finishing process, and the material's tensile strength. For steels, $K_\mathrm{s}$ is frequently determined from the fatigue endurance limit, which typically corresponds to the fatigue strength at $10^6$ cycles. This definition provides a reference point and highlights the importance of surface roughness in the high-cycle fatigue (HCF) regime, as extensively observed \cite{Maiya1975,Bayoumi1995, Fordham1997, Murakami2002, Arola2002, Itoga2003, Suraratchai2008, Singh2019, Vayssette2019,Gustafsson2024}. However, identifying an accurate methodology to define $K_\mathrm{s}$ remains challenging due to the variety of alloys and machining processes \cite{As2005}. 

Several studies have focused on deriving expressions for the surface factor based on experimental fatigue data. In 1946, Noll and Lipson \cite{Noll1946} compiled an extensive dataset on the fatigue strength of steels across a range of hardness values and surface conditions. Their findings have been widely applied \cite{Juvinall2020} and contributed to the development of empirical formulas for steel subjected to different machining processes \cite{Shigleys2011}. In 1973, Johnson \cite{Johnson1973} introduced a diagram correlating $K_\mathrm{s}$ with both surface metrics and tensile strength for machined and ground steel surfaces, highlighting that the effect of surface roughness is stronger for higher values of tensile strength and average surface roughness. A standardised definition of $K_\mathrm{s}$ was later published in the 1994 FKM guideline \cite{Rennert2024}. Since then, further research has explored the interplay between surface roughness and fatigue. In 2003, Itoga et al. \cite{Itoga2003} investigated the changes in the endurance limit of very high-strength steels under different surface roughness conditions using rotary bending fatigue tests. Through experimental studies, Deng et al. \cite{Deng2009} concluded that surface roughness has a more pronounced effect on crack initiation than on crack propagation. Although these studies provide a first approximation, the resulting empirical approaches are often too conservative or limited \cite{Singh2019}. McKelvey and Fatemi \cite{McKelvey2012} analysed the combined effects of surface finish and hardness on the fatigue behavior of forged steel, proposing some modifications with respect to previous studies \cite{Noll1946}. Shareef and Hasselbusch \cite{Shareef1996} highlighted that the dataset provided in Ref. \cite{Noll1946} is limited to a maximum hardness of 33 HRC. All of the aforementioned analyses primarily focus on steel.  In contrast, Sinclair et al. \cite{Sinclair1957} reported that the surface roughness has a minimal impact on the fatigue life of titanium alloys. Moreover, the growing use of additively manufactured metals has prompted numerous experimental studies, emphasizing surface roughness as a key factor in the fatigue behavior of these materials \cite{Solberg2019, Lee2020, Zhang2019}.

A number of studies have modelled surface roughness as a series of microscopic notches, deriving expressions to estimate the surface factor with improved accuracy. Initially, a static stress concentration factor $K_t$ was introduced in 1961 by Neuber based on fracture mechanics principles \cite{Neuber1961}. However, since $K_t$ is known to provide very conservative results, it was modified through the application of semi-empirical expressions, such as the one suggested by Peterson \cite{Arola2002,Peterson1959}. This approach has been utilised in a semi-analytical framework \cite{Arola2002} or in conjunction with finite element analysis of the measured surface topography \cite{Suraratchai2008,As2005}.  However, although these methodologies are highly accurate in estimating $K_t$, their reliability in estimating the impact on fatigue relies on empirical approximations that fail to account for the wide variety of surface finishes and materials influenced by this phenomenon. In 1983, Murakami et al. \cite{Murakami1983} introduced a model to predict the fatigue strength of components with small defects, based on geometrical parameters. Despite its limitations, this model has been successfully applied to investigate the influence of surface roughness on fatigue performance \cite{Murakami2002,Murakami1994}. Moreover, Vayssette et al. \cite{Vayssette2019} applied the Crossland criterion to estimate the endurance limit of rough metallic surfaces. However, this expression relies on additional parameters that need to be experimentally calibrated \cite{Crossland1956}.  

An alternative to these semi-analytical techniques is the application of a continuum damage model (CDM) \cite{Chaboche1989} combined with a topological representation of the material's microstructure. In this kind of models, widely employed to predict crack initiation and propagation \cite{Li2023}, a damage variable is introduced to describe the material degradation. The extension of CDMs to the analysis of rough surfaces has been applied to various phenomena, including fretting wear of rough Hertzian contacts \cite{Leonard2013}, crack nucleation in line contacts \cite{Agham2014}, tensile tests on steel with low average surface roughness \cite{Singh2019} and rolling contact fatigue \cite{Lorentz2021}. However local CDMs are known to be mesh-dependent due to the loss of ellipticity of the governing equations \cite{Li2023,Peerlings2000}. This issue can be addressed by introducing a gradient term into the formulation, as demonstrated in the well-known phase field model for fracture \cite{Marigo2016, Bourdin2007}. Phase field fracture formulations can be extended to fatigue by introducing a fatigue degradation function that diminishes the fracture energy over cycles, depending on a cyclic history variable \cite{Alessi2018, Carrara2020}. This concept was extended to the HCF regime by Golahmar et al. \cite{Golahmar2023}, and it has been experimentally verified and successfully applied in the literature; for example, to predict hydrogen-assisted fatigue crack growth \cite{Cui2024}. Despite its applicability, the phase field fracture model has not been applied to study the effect of surface roughness in fatigue failure, although an initial application for static fracture in adhesive contact problems with embedded roughness was succesfully assessed \cite{Paggi2020}.

As mentioned above, a key aspect in the numerical study of surface roughness is the design of a mesh that accurately reproduces the surface topology, which is inherently stochastic \cite{Loth2023}. Several approaches have been proposed in the literature. The simplest technique involves a sinusoidal pattern, requiring real surface measurements to ensure accuracy \cite{Lorentz2021}. Some researchers have applied optical profilometers (e.g. a white light interferometer \cite{As2005}) to measure the surface topology and use these measurements to construct the mesh \cite{Vayssette2019, As2005}. However, this approach is limited by the small number of specimens analysed and does not account for the stochastic nature of roughness. An alternative approach considers microstructures generated by Voronoi tesselations \cite{Singh2019, Leonard2013}, with randomly extruded nodes defining the roughness profile. However, this method does not capture the correlation length of real profiles \cite{Loth2023} and depends on average roughness. Also, other metrics such as root mean square roughness, are more effective in fatigue analysis, as they better represent the most critical defects found on the workpiece surface \cite{Novovic2004}. Moreover, accurate results are obtained only if the grain size is known \cite{Singh2019}, once again reliying on optical measurements (planimetric method). Recently, Loth et al. \cite{Loth2023} introduced a novel methodology for generating rough profiles in 3D surfaces. The tool has been successfully applied in optics \cite{Loth2023}, and it is based on a stochastic distribution that accounts for the correlation length of the surface profile. 

In this context, the main objective of this paper is to develop a novel numerical framework able to describe the effect of surface roughness, especially in the HCF regime. The proposed model builds upon the state-of-the-art phase field approach for fatigue fracture introduced in Ref. \cite{Golahmar2023}, and integrates it with the stochastic meshing strategy for rough surfaces presented in Ref. \cite{Loth2023}. The key novel contribution of this work lies in the integration of these two methodologies into a unified numerical framework that enables a direct and quantitative analysis of the relationship between surface topography, material properties, and fatigue life. To the best of the author's knowledge, our work is the first to numerically establish a relationship between metrics of surface roughness, material properties and fatigue life (surface factor). This brings new insight into how much the fatigue life is reduced for a given surface roughness and material properties, without the need for extensive and costly experimental campaigns. The paper is divided as follows. Section \ref{sec2} describes the numerical model developed. Section \ref{sec3} presents the results obtained. The first part verifies the newly developed computational framework against experimental data from the literature, while the second presents the evolution of the surface factor with respect to key parameters. Section \ref{sec4} outlines the main conclusions of this study. Finally, \ref{app1} provides further details on the process for determining the surface factor and the computational cost of the tool.

\section{A computational model for predicting fatigue life of rough specimens}\label{sec2}

The numerical framework presented in this section integrates a phase field model for fatigue fracture with a meshing tool designed for rough specimens. The fracture model, detailed in Section \ref{sec2a}, includes an additional function within the phase field formulation to characterise the degradation of fracture toughness over loading cycles. Meanwhile, the roughness-generation tool, described in Section \ref{sec2b}, accounts for the stochastic nature of surface roughness, enabling the specimen geometry to be defined based on commonly used roughness profile metrics.

\subsection{A phase field model for fatigue fracture}\label{sec2a}

In the phase field model for fracture, the crack is represented by a smooth, continuous scalar variable $\phi$, referred to as the phase field, which is akin to a damage variable, and ranges between $0$ (intact material) and $1$ (fully broken material). A key condition of the model is the monotonic growth of $\phi$, expressed as $\dot{\phi} \geq 0$. Introducing a damage variable makes the problem computationally tractable. It allows the work to create a crack interface $\Gamma$ to be expressed as a volume integral in the variational form of Griffith’s criterion, instead of a surface integral \cite{Bourdin2000}. Hence, in a volume $\Omega \in \mathbb{R}^\delta$ ($\delta \in {1, 2, 3}$), bounded by the external surface $\partial \Omega \in \mathbb{R}^{\delta-1}$, with the outward normal vector $\mathbf{n}$,
\begin{equation}
   \int_{\Gamma} G_\mathrm{c} \, \mathrm{d}S \approx \int_{\Omega} \psi_f(\phi, \nabla\phi) \, \mathrm{d}V, \label{eq:fracture_energy}
\end{equation}
being $\psi_f(\phi, \nabla\phi)$ the fracture energy density. Consider that $\Omega$ is subjected to a cyclic body force $\mathbf{b}$, a cyclic traction per unit area $\mathbf{t}$ on $\partial_{t} \Omega$, and a cyclic imposed displacement $\mathbf{\bar{u}}$ on $\partial_u \Omega$, such that $\partial \Omega = \partial_{t} \Omega \cup \partial_u \Omega$. The solution $(\mathbf{u}, \phi)$, where $\mathbf{u}$ represents the displacement field of the solid, is governed by the principle of energy balance, which is satisfied when the internal power of the system is equal to the external power \cite{Carrara2020}. This can be written as follows, 
\begin{equation}
   \int_{\Omega} \dot{\psi} (\boldsymbol{\varepsilon}(\mathbf{\mathrm{u}}),\phi,\nabla \phi \mid \bar{\alpha}) \, \mathrm{d}V - \int_{\partial_{t} \Omega} \mathbf{t} \cdot \dot{\mathbf{u}} \, \mathrm{d}S - \int_{\Omega} \mathbf{b} \cdot \dot{\mathbf{u}} \, \mathrm{d}V = 0,  \label{eq:balance_power}
\end{equation}
with the vertical line used to distinguish between the current state variables ($\mathbf{u}, \phi, \nabla\phi$) and the cumulated fatigue history variable $\bar{\alpha}$. This notation emphasizes that $\bar{\alpha}$ is treated as a known parameter during the evaluation of Eq. \eqref{eq:balance_power}, while the other quantities are considered as the primary variables. In Eq. \eqref{eq:balance_power}, the internal energy density of the solid, $\psi(\boldsymbol{\varepsilon}(\mathbf{u}), \phi, \nabla \phi \mid \bar{\alpha})$, is defined as the sum of two distinct terms: the elastic energy density, $\psi_{\mathrm{e}}(\phi, \boldsymbol{\varepsilon}(\mathbf{u}))$, and the fracture energy density, $\psi_{\mathrm{f}}(\phi, \nabla \phi \mid \bar{\alpha})$,
\begin{equation}
    \psi(\boldsymbol{\varepsilon}(\mathbf{u}),\phi,\nabla \phi \mid \bar{\alpha}) = \psi_{\mathrm{e}}(\phi, \boldsymbol{\varepsilon}(\mathbf{u})) + \psi_{\mathrm{f}}(\phi, \nabla \phi \mid \bar{\alpha}), \label{eq:energydensity}
\end{equation}
where $\boldsymbol{\varepsilon}(\mathbf{u})$ is the strain tensor, that is defined assuming small deformations,
\begin{equation}
    \boldsymbol{\varepsilon}(\mathbf{u}) = \frac{1}{2} \left( \nabla \mathbf{u} + \nabla^\mathrm{T} \mathbf{u} \right). \label{eq:strain_tensor}
\end{equation}
The elastic energy density, $\psi_{\mathrm{e}}(\phi, \boldsymbol{\varepsilon}(\mathbf{u}))$, represents the stored energy in the solid. It is divided into two components \cite{Miehe2010a} to prevent compressive strains from contributing to crack nucleation or propagation,
\begin{equation}
    \psi_{\mathrm{e}}(\phi, \boldsymbol{\varepsilon}(\mathbf{u})) = g(\phi) \psi_{e0}^+(\boldsymbol{\varepsilon}(\mathbf{u})) + \psi_{e0}^-(\boldsymbol{\varepsilon}(\mathbf{u})), \label{eq:energysplit}
\end{equation}
where the index $0$ denotes the undamaged state of the elastic energy. Various energy splits have been proposed in the literature \cite{Golahmar2023}. The one applied in this paper is the no-tension energy split introduced in Ref. \cite{Freddi2010}, originally developed for masonry-like materials, as it provides a detailed and physically-consistent modelling of metallic fatigue in the HCF regime \cite{Golahmar2023}. Thus, the no-tension split is expressed as in Ref. \cite{Lo2019},
\begin{eqnarray}
    \psi_\text{e0}^+ =
    \begin{cases} 
    \text{if} \quad \varepsilon_{III}>0, \quad \frac{\lambda}{2} \left( \varepsilon_{I} + \varepsilon_{II} + \varepsilon_{III} \right)^2 + \mu \left( \varepsilon_{I}^2 + \varepsilon_{II}^2 + \varepsilon_{III}^2 \right) , \\
    \text{else if} \quad \varepsilon_{II} + \nu \varepsilon_{III}>0, \quad \frac{\lambda}{2} \left( \varepsilon_{I} + \varepsilon_{II} + 2\nu \varepsilon_{III} \right)^2 + \mu \left( (\varepsilon_{I}+\nu \varepsilon_{III})^2 + (\varepsilon_{II}+\nu \varepsilon_{III})^2 \right), \\
    \text{else if} \quad (1-\nu)\varepsilon_{I} + \nu (\varepsilon_{II}+\varepsilon_{III})>0, \quad \displaystyle \frac{\lambda}{2\nu(1-\nu)} \left( (1-\nu)\varepsilon_{I} + \nu (\varepsilon_{II}+\varepsilon_{III}) \right), \\
    \text{else} \quad 0.
    \end{cases}, \\
    \text{and} \qquad \psi_\text{e0}^- =
    \begin{cases} 
    \text{if} \quad \varepsilon_{III}>0, \quad 0 , \\
    \text{else if} \quad \varepsilon_{II} + \nu \varepsilon_{III}>0, \quad \frac{E}{2} \varepsilon_{III}^2 \\
    \text{else if} \quad (1-\nu)\varepsilon_{I} + \nu (\varepsilon_{II}+\varepsilon_{III})>0, \quad \displaystyle \frac{E}{2(1-\nu^2)} \left( \varepsilon_{II}^2 + \varepsilon_{III}^2 + 2\nu \varepsilon_{II}\varepsilon_{III} \right), \\
    \text{else} \quad \frac{\lambda}{2} \left( \varepsilon_{I} + \varepsilon_{II} + \varepsilon_{III} \right)^2 + \mu \left( \varepsilon_{I}^2 + \varepsilon_{II}^2 + \varepsilon_{III}^2 \right),
    \end{cases}
\end{eqnarray}

\vspace{0.2cm}
\noindent being $\varepsilon_{III} \leq \varepsilon_{II} \leq \varepsilon_{I}$ the principal strains, ($\lambda, \mu$) the Lamé coefficients, and $\nu$ denoting Poisson's ratio. 

In Eq. \eqref{eq:energysplit} the function $g(\phi)$ represents the degradation function. This function must satisfy the following conditions:
\begin{equation}
	g(0) = 1, \quad g(1) = 0, \quad \mathrm{and} \quad g'(\phi) \leq 0. \label{eq:degfunc}
\end{equation}
In this paper, a quadratic degradation function 
\begin{equation}
	g(\phi) = (1-\phi)^2 \label{eq:degfunc_def}
\end{equation}
is selected, as it is commonly established in the literature \cite{Carrara2020, Golahmar2023}.

On the other hand, the fracture energy density $\psi_{\mathrm{f}}(\phi, \nabla \phi \mid \bar{\alpha})$ is defined as 
\begin{equation}
	\psi_{\mathrm{f}}(\phi, \nabla \phi \mid \bar{\alpha}) = f(\bar{\alpha}) \frac{G_\text{c}}{4 c_w} \left(\frac{w(\phi)}{\ell} + \ell ||\nabla \phi||^2  \right), \label{eq:fracene}
\end{equation}
where $G_\text{c}$ is the critical energy release rate, the function $w(\phi)$ is the dissipation function, which satisfies $w(0) = 0$ and $w(1) = 1$, and the coefficient $c_w$ is defined as $c_w = \int_0^1 \sqrt{w(s)} \, \mathrm{d}s$. Various definitions of $w(\phi)$ and $c_w$ have been proposed in the literature. One of the most widely used ones is the AT1 model \cite{Pham2011}, which assumes an initial elastic phase with no damage ($\phi = 0$), followed by a damaged phase where $\phi$ evolves until complete damage ($\phi = 1$),
\begin{equation}
	w(\phi) = \phi \quad \text{and} \quad c_w = 2/3. \label{eq:at1model}
\end{equation}
In this model, the so-called phase field length scale $\ell$ can be defined as a function of Irwin's characteristic length $\ell_\text{Irwin}$, obtaining
\begin{equation}
    \ell = \frac{3}{8} \ell_\text{Irwin} = \frac{3 E G_\text{c}}{8\sigma_\text{c}^2} \label{eq:AT1_ell}
\end{equation} 
where $E$ is the Young's modulus and $\sigma_\mathrm{c}$ the fracture strength \cite{Mandal2024}. 

There are several definitions in the literature for the fatigue degradation function $f(\bar{\alpha})$ \cite{Carrara2020}, which characterises the reduction in fracture toughness as a function of the cumulated fatigue history $\bar{\alpha}$. In this paper, the so-called $f_2(\bar{\alpha})$ function is employed \cite{Golahmar2023},
\begin{equation}
    f_2(\bar{\alpha}) = \left( 1 - \frac{\bar{\alpha}}{\bar{\alpha}_{\mathrm{T}}} \right)^2, \label{eq:degfunc}
\end{equation}
where $\bar{\alpha}_{\mathrm{T}}$ is a material parameter that can be obtained through calibration with experiments. An estimation of this parameter is defined considering a point of the S-N curve ($N_\square, \sigma_\square$). This point preferably corresponds to low stress amplitudes, where the macroscopic behaviour is linear elastic and the S-N curve exhibits a single slope \cite{Golahmar2023},
\begin{equation}
    \bar{\alpha}_{\mathrm{T}} = \frac{N_\square \displaystyle{} \left( \frac{\sigma_\square}{\sigma_\text{c}} \right)^{2n}}{\left( 1 - \displaystyle{} \frac{\sigma_\square}{\sigma_\text{c}} \right)}. \label{eq:a0def}
\end{equation}
The exponent $n$ proposed in Ref. \cite{Golahmar2023} depends on the slope of the S-N curve, which can be approximated using Basquin's law $\sigma = a N^{-b}$. 
\begin{equation}
    n = C_1 (1/b) + C_2, \label{eq:param_n}
\end{equation}
where the coefficients $C_1$ and $C_2$ are calibrated in Ref. \cite{Golahmar2023} for different degradation functions and phase field approaches. In this paper $C_1=0.5$ and $C_2=-0.13$ for the $f_2(\bar{\alpha})$ degradation function and the AT1 model.

The so-called cumulated fatigue history variable $\bar{\alpha}$ is defined incrementally. Hence, for a time step $i$, 
\begin{equation}
    \bar{\alpha}_i = \bar{\alpha}_{i-1} + \Delta \bar{\alpha}, \label{eq:fathis}
\end{equation}
considering that $\bar{\alpha}_0=0$. Then, the increment in the cumulated fatigue history variable $\Delta \bar{\alpha}$ is defined following Ref. \cite{Golahmar2023}, to account for (i) the slope of the S-N curve, (ii) the endurance limit and (iii) the effect of the stress ratio $R$ between the minimum and the maximum stress in the loading cycle ($\sigma_{\mathrm{min}}/\sigma_{\mathrm{max}}$),
\begin{equation}
    \Delta \bar{\alpha} = \left( \frac{\alpha_\text{max} (1-R)}{2\alpha_\text{n}}\right)^{n} \text{Heaviside}\left( \max_{\tau \in [0,t]} \left( \frac{\alpha_\text{max}(1-R)}{2} \right) - \alpha_e \right). \label{eq:fathisincr}
\end{equation}
In Eq. \eqref{eq:fathisincr} the parameter $\alpha_n$ is defined assuming the AT1 problem, $\alpha_n=\frac{\sigma_{{\mathrm{c}}}^2}{2E}$, and used to add dimensional consistency. The energy endurance in the S-N curve $\alpha_e$ is defined as a local variable $\alpha_e = \psi_\text{e0}^+(\boldsymbol{\varepsilon}(\mathbf{u_\text{e}}))$, where $\mathbf{u_\text{e}}$ represents the displacement field in the solid when the stress amplitude of the loading cycle equals the endurance limit $\sigma_e$. An initial approximation of this parameter can be computed by assuming a 1D problem, $\alpha_e = \frac{\sigma_{\mathrm{e}}^2}{2E}$. However, this initial value might be too conservative for geometries with significant stress concentrators. Finally, $\alpha_\mathrm{max}$ is defined as the maximum fatigue history in the loading cycle, corresponding to the maximum damaged positive strain energy density stored in the solid during the cycle: $\alpha_\mathrm{max} = g(\phi) \psi_\text{e0}^+(\boldsymbol{\varepsilon}(\mathbf{u}))$ \cite{Carrara2020, Golahmar2023}.

Applying standard arguments of variational calculus in Eq. \eqref{eq:balance_power}, the following mechanical problem is obtained
\begin{eqnarray}
    \nabla \cdot \boldsymbol{\sigma} + \mathbf{b} &=& \boldsymbol{0} \qquad \text{in $\Omega$}, \label{eq:strong_eq}\\
    \boldsymbol{\sigma} \cdot \mathbf{n} &=& \mathbf{t} \qquad \text{on $\partial_{t} \Omega$}, \label{eq:strong_cdc}
\end{eqnarray} 
toghether with the displacement boundary conditions and the following damage problem
\begin{equation}
    g'(\phi) \mathcal{H} + \frac{G_\mathrm{c} f(\bar{\alpha})}{ 2 c_w } \left( \frac{w'(\phi)}{2\ell} - \ell \nabla^2 \phi \right) - \frac{G_\mathrm{c} \ell}{2c_w} \nabla \phi \, \cdot \, \nabla f(\bar{\alpha})  = 0 \qquad \text{in $\Omega$},
\end{equation} 
where $\mathcal{H}$ is the so-called history variable, that is defined as 
\begin{equation}
    \mathcal{H} = \mathrm{max}\left\{ \max_{\tau \in [0,t]} \psi_\text{e0}^+(\boldsymbol{\varepsilon}(\mathbf{u})), \mathcal{H}_\mathrm{min}  \right\}, 
\end{equation} 
being $\mathcal{H}_\mathrm{min}=\frac{3G_\mathrm{c}}{16\ell} $ for the AT1 model. Note that in this formulation, the energy split is applied only to the damage evolution problem, the so-called hybrid approach \cite{Ambati2015}.  The numerical application of the PF model is achieved using the Finite Element Method (FEM) and a staggered scheme. At each time step $t$, an iterative procedure is initiated. First, the displacement field $\mathbf{u^i}$ at iteration $i$ is computed while keeping the damage variable fixed, i.e., $\phi^i = \phi^{i-1}$. Subsequently, the damage variable $\phi^i$ is updated while holding the displacement field $\mathbf{u^i}$ constant. This process is repeated until convergence is achieved, which is determined by the condition $|\phi^i - \phi^{i-1}| < \text{Tol}$, where $\text{Tol}$ is a tolerance that must be defined according to the specific problem (in this paper, $\text{Tol} = 10^{-6}$). It should be noted that the first iteration, $i=1$, uses the damage variable from the previous time step $t-1$.

\subsection{A stochastic tool for generating numerical surface roughness}\label{sec2b}

\begin{figure} [H]
     \centering
     \begin{subfigure}[t]{0.58\textwidth}
         \centering
         \includegraphics[width=\textwidth]{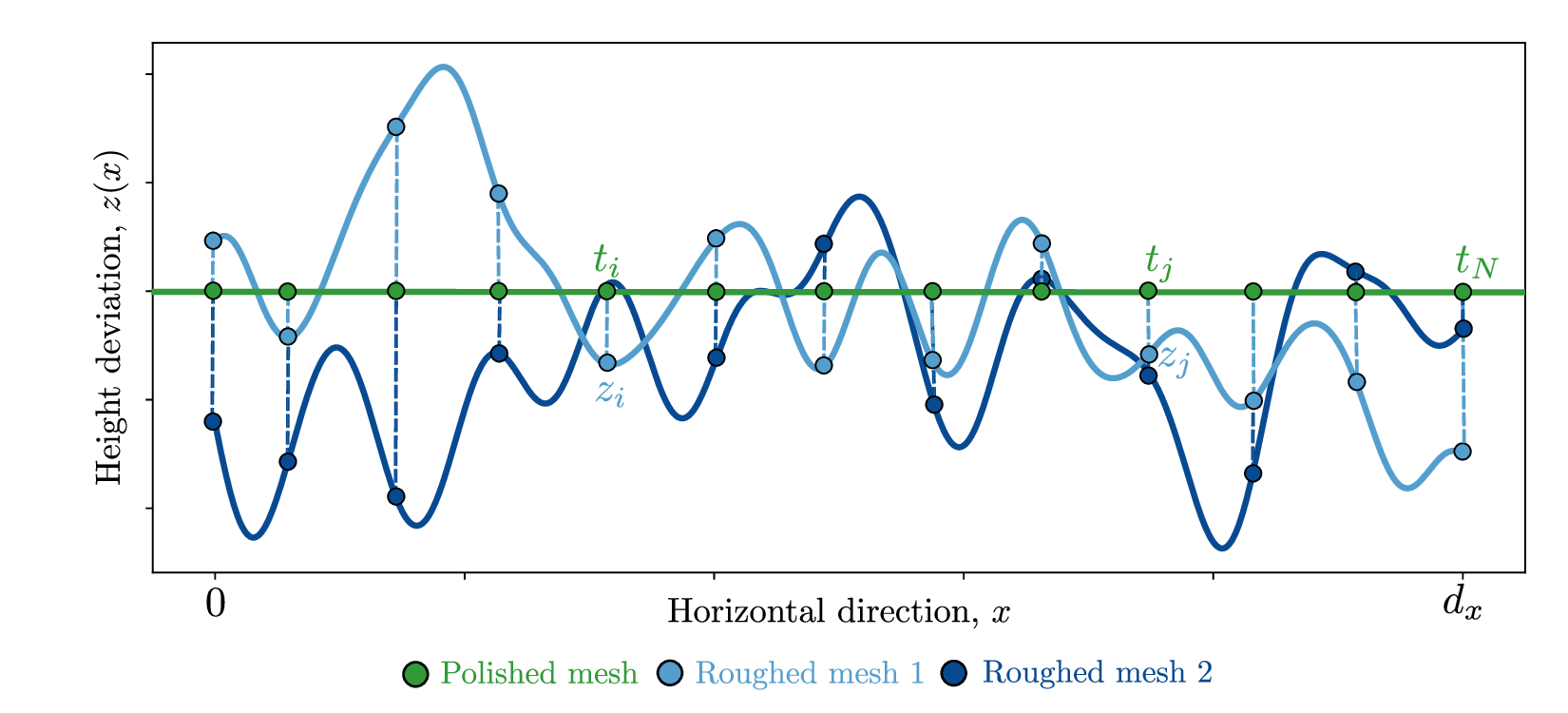}
         \caption{Surface profile}
         \label{fig:1Dprofile}
     \end{subfigure}
     \hfill
	\begin{subfigure}[t]{0.39\textwidth}
         \centering
         \includegraphics[width=\textwidth]{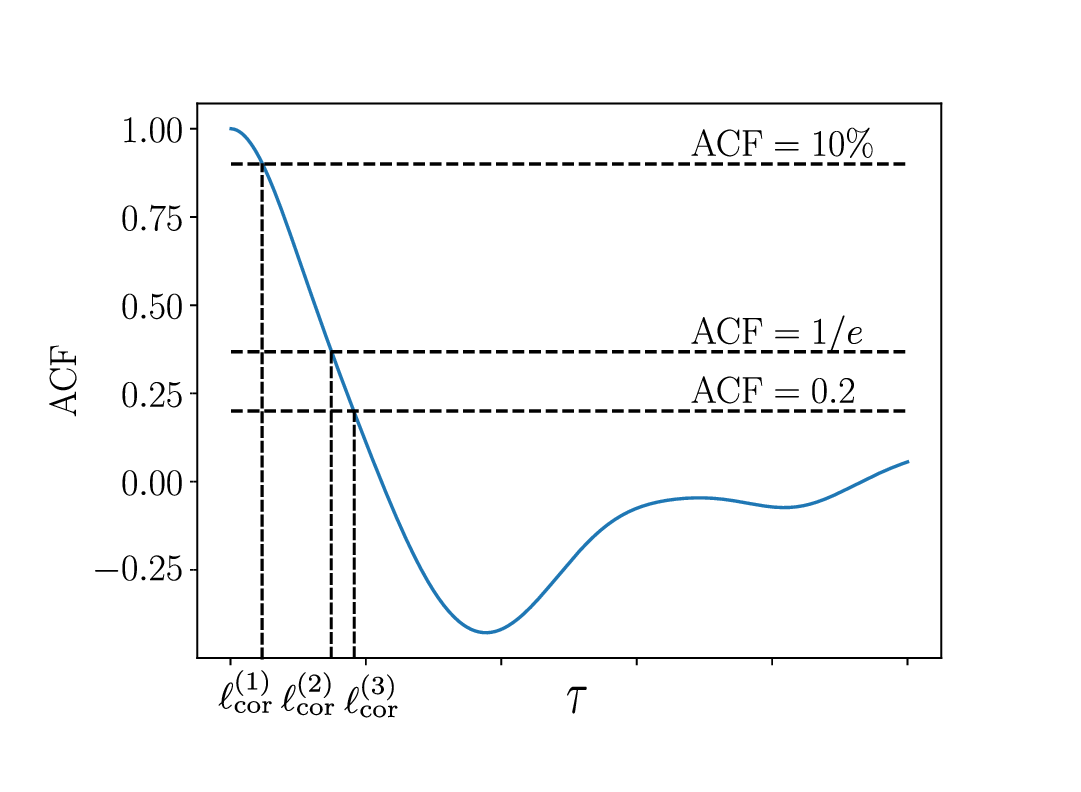}
         \caption{ACF}
         \label{fig:ACF}
    \end{subfigure}
    \caption{Analysis of the rough surface topology in a 1D profile. In (a) two examples are represented in blue colours. The green points represent the smooth surface. In (b) the autocorrelation function (ACF) is represented, highlighting the three criteria proposed for determining the correlation length.}
        \label{fig:topology}
\end{figure}

Surface roughness refers to the deviations in the normal direction $z$ of a real surface from its nominal form. It is commonly characterised using various metrics, with the average surface roughness ($R_a$) and the root mean square roughness ($R_q$) being the most widely used. The definitions of these parameters are frequently represented in a discrete form for a given sample of $N$ points on a real surface \cite{Gadelmawla2002}:
\begin{equation}
    R_a = \frac{1}{N} \sum_{i=1}^N |z_i|, \qquad \text{and} \qquad R_q = \sqrt{ \frac{1}{N} \sum_{i=1}^N z_i^2}.
\end{equation}
In the model introduced in Ref. \cite{Loth2023}, the rough profile is obtained deviating a given set of $N$ points on the mesh of the smooth (polished) surface (see the green points of Fig. \ref{fig:1Dprofile}). The deviation vector $\mathbf{z}$ is defined following a Gaussian stochastic procedure:
\begin{equation}
    \mathbf{z} = [z_0, ..., z_i, ..., z_N] = R_q \boldsymbol{L} \mathbf{w}, \label{eq:zmesh}
\end{equation}
where $\mathbf{w}$ is a Gaussian white noise variable (a vector of size N) that is calculated using a random number generator. This generates different rough profiles, as the ones highlighted in the blue points of Fig. \ref{fig:1Dprofile}. The matrix $\boldsymbol{L}$ (size NxN) is obtained from the Cholesky decomposition of the autocorrelation matrix in the distribution $\boldsymbol{R} = \boldsymbol{L} \boldsymbol{L}^T$. Each term of $\boldsymbol{R}$ is calculated as
\begin{equation}
    R_{ij} = \exp\left( - \frac{\left| x_{0i} - x_{0j} \right|^2}{2\ell_\mathrm{cor}^2} \right), \label{eq:Rmatrix}
\end{equation}
being $\mathbf{x}_0$ the initial position of the nodes in the mesh of the polished surface. The parameter $\ell_\mathrm{cor}$ denotes the correlation length that represents the characteristic spacing between surface features. It is determined as a certain threshold $\tau$ of the autocorrelation function (ACF) using various criteria. The most notable ones found in the literature are represented in Fig. \ref{fig:ACF} and include: (1) the value of $\tau$ at which the ACF decreases to $10\%$ of its initial value \cite{Gathimba2019}, (2) the value of $\tau$ at which the ACF equals $1/e$ \cite{Maradudin1989}, where $e$ is the Napier constant \cite{Muralikrishnan2009}, and (3) the value of $\tau$ at which the ACF reaches $0.2$ \cite{Blateyron2013}. In this paper, these options are analysed to identify the most suitable criterion. The ACF for a one-dimensional profile sample of length $d_x$ is written as
\begin{equation}
    \mathrm{ACF}(\tau) = \frac{\int_0^{d_x} z(x) z(x-\tau) dx}{\int_0^{d_x} z^2(x) dx}, \label{eq:ACF}
\end{equation}
where $d_x$ depends on the roughness level, as different levels require varying sampling resolutions. Reference threshold values are specified in the ISO/FDIS 2190-3 standard \cite{ISO2021}. For instance, $d_x =$ \SI{500}{\micro\meter} is recommended for $R_a$ values between \SIrange{0.1}{2}{\micro\meter}. To minimise stochastic errors, multiple ACF measurements are typically performed. Standards also specify the required sample size for each roughness level, defining a length $L_e$ over which $d_x$ should be measured as many times as possible \cite{ISO2021}. 

Note that, since surface roughness follows a Gaussian distribution, either $R_a$ or $R_q$ can be used as inputs for the tool, as they are related by $R_q\approx 1.25 R_a$ \cite{Raoufi2015}, highlighting the versatility of this numerical tool.

\begin{figure}[h]
    \centering
    \includegraphics[scale=1.0]{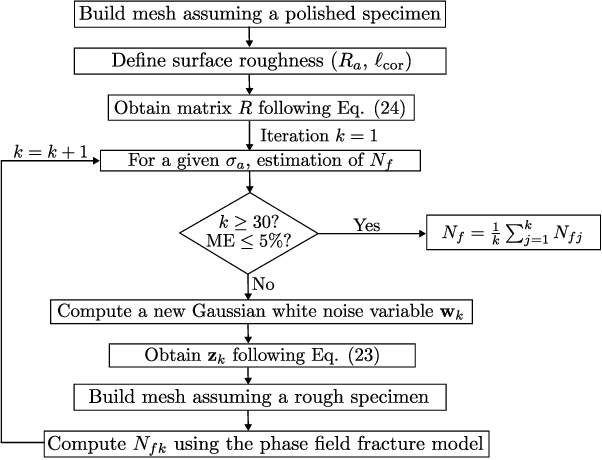}
    \caption{Computational framework flowchart for determining the number of cycles to failure ($N_f$) of a specific rough specimen under a given stress amplitude $\sigma_a$, with characteristic roughness parameters $R_a$ and $\ell_\mathrm{cor}$.}\label{fig:flowchart}
\end{figure}

The stochastic nature of the mesh implies that the results should be analysed based on stochastic principles. First, although the mesh topology is designed to follow a Gaussian distribution, this does not necessarily apply to the phase field results. To address this, at least $M=30$ samples are analysed in each case, ensuring a sufficient sample size to support the assumption of a Gaussian distribution according to the Central Limit Theorem \cite{Montgomery2011}. Second, a minimum confidence interval of $95\%$ is established as a criterion for determining the sample size. Accordingly, a relative margin of error ($\mathrm{ME}$) is calculated using the fatigue life obtained for each of the rough profiles in the sample $\boldsymbol{N}_f = [N_{f0}, ..., N_{fk}, ..., N_{fM}]$, 
\begin{equation}
    \mathrm{ME} [\%] = \frac{1.96 \, s(\boldsymbol{N}_f)}{\sqrt{M}} N_f  \cdot 100, \label{eq:me}
\end{equation}
where $N_f$ is the expected value of $\boldsymbol{N}_f$ and $s(\boldsymbol{N}_f)$ is the standard deviation of $\boldsymbol{N}_f$. In this paper, the sample size is estimated such that $M\geq 30$ and $\mathrm{ME} \leq 5\%$.  Thus, Fig. \ref{fig:flowchart} shows the flowchart for calculating, for a given stress amplitude $\sigma_a$, the number of cycles to failure $N_f$ of a rough specimen with roughness metrics $R_a$ and $\ell_\mathrm{cor}$. Initially, the mesh of the specimen is defined by assuming its surface is smooth, without roughness. This step enables the definition of vector $\mathbf{x}_{0}$, and therefore the calculation of matrix $\mathbf{R}$ applying Eq. \eqref{eq:Rmatrix}. Subsequently, an iterative process is performed, which concludes when a minimum of 30 iterations is reached, and the $\mathrm{ME}\leq 5\%$ condition is satisfied. At each iteration $k$, the vector $\mathbf{z}_k$ is obtained applying a different white noise vector $\mathbf{w}_k$ as described in Eq. \eqref{eq:zmesh}, which creates a rough mesh for the specimen. Then, the phase field fracture model described in Section \ref{sec2a} is applied to obtain the value of $N_{fk}$. Finally, the mean of the $N_{fk}$ values, denoted as $N_{f}$, is calculated. It is important to note that the results obtained from this numerical framework are therefore inherently non-deterministic. 

The numerical framework developed in this work has been implemented using FEniCSx, an open-source FEM software \cite{fenicsx}, and uses also Gmsh, an open-source 3D finite element mesh generator \cite{gmsh}. The codes developed are made freely available to download at https://mechmat.web.ox.ac.uk/codes

\section{Results}\label{sec3}

\begin{figure}[H]
    \centering
    \includegraphics[scale=0.6]{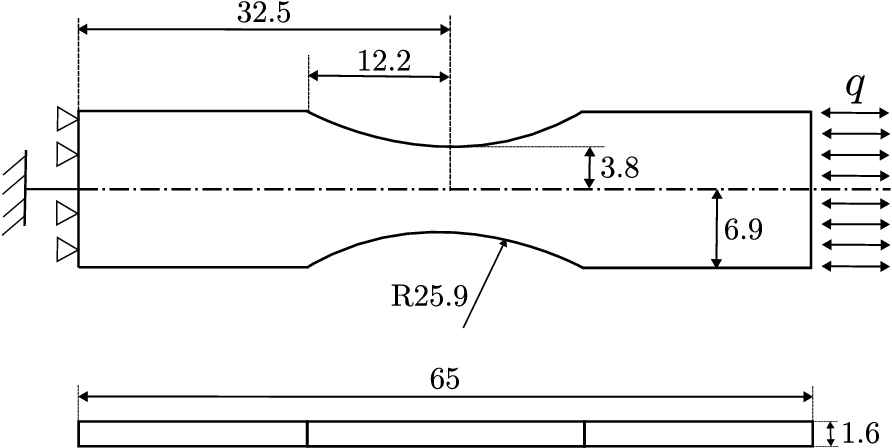}
    \caption{Geometry of the problem. Dimensions, in mm, taken from Ref. \cite{Singh2019}. The applied distributed force is denoted as $q$.}\label{fig:modelo}
\end{figure}

This section is divided into two parts. In Section \ref{sec3a} the results provided by the model are compared with experimental data from the literature. In Section \ref{sec3b} the correlation between the surface factor and key parameters is established. Throughout the analysis, the boundary value problem shown in Fig. \ref{fig:modelo} is used as in Ref. \cite{Singh2019}. The experiments involve tension-compression fatigue with a load ratio $R=-1$. Due to the small thickness of the specimen, plane stress is assumed.

The material adopted is AISI 4130, and its mechanical properties are summarised in Table \ref{table:mechprop}. The Young’s modulus ($E$), Poisson’s ratio ($\nu$), and ultimate strength ($\sigma_\mathrm{ULT}$) are sourced from Ref. \cite{Singh2019}, while the fracture toughness ($K_\mathrm{IC}$) is obtained from Ref. \cite{Lai1975}. Note that the critical energy release rate $G_\mathrm{c}$, as used in Eqs. \eqref{eq:fracene} and \eqref{eq:AT1_ell}, is related to  $K_\mathrm{IC}$ following the well-stablished relation $G_\mathrm{c} = K^2_\mathrm{IC} / E$ under plane stress conditions.

\renewcommand{\arraystretch}{1.5}
\begin{table}[H]
    \centering
    \begin{tabular}{c c c c c c c c c}
    \hline
    $E$ [GPa] & $\nu$ & $\sigma_\mathrm{ULT}$ [MPa]  & $K_\mathrm{IC}$ [MPa $\sqrt{\mathrm{m}}$] & $G_\mathrm{c}$ [MPa m] & $\sigma_\mathrm{c}$ [MPa] & $\ell$ [mm] & $\sigma_\mathrm{e}$ [MPa] & $\bar{\alpha}_{\mathrm{T}}$ \\ 
    \hline
    200 & 0.3 & 530 & 60.5 & 0.018 & 1121 & $2.9$ & 263 & $8.2 \cdot 10^{-4}$\\
    \hline
    \end{tabular}
    \caption{Mechanical properties estimated for the AISI 4130 steel.}\label{table:mechprop}
\end{table}

\begin{figure}[H]
    \centering
    \includegraphics[scale=0.8]{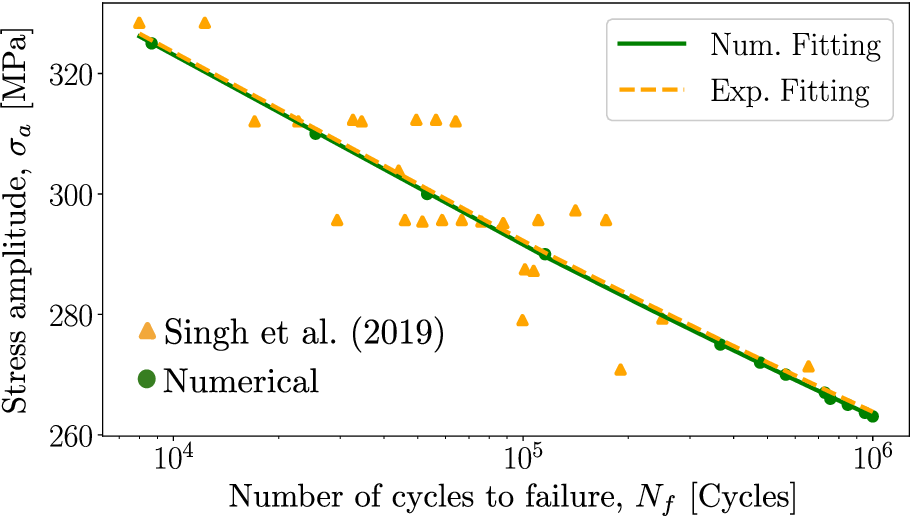}
    \caption{S-N curve of the polished surface. Comparison with experiments \cite{Singh2019}. }\label{fig:SNpolished}
\end{figure}

For brittle materials, the fracture strength, also named as tensile strength ($\sigma_\mathrm{c}$) directly corresponds to $\sigma_\mathrm{ULT}$. In metals, however, $\sigma_\mathrm{ULT}$ is typically defined as the maximum engineering stress observed during testing, which coincides with the onset of significant reduction in the specimen’s cross-sectional area. This parameter does not represent the true fracture stress, which is higher than $\sigma_\mathrm{ULT}$ and is often unavailable due to the complexities in accurately adjusting the triaxial stress state at fracture in tensile specimens \cite{Dieter1976}. To address these limitations, the magnitude of $\sigma_\mathrm{c}$ reported in Table \ref{table:mechprop} has been adjusted to align with the experimental S-N curve determined for the polished specimen \cite{Singh2019}, where the tensile stress at the specimen's inner surface ($\sigma$) is related to the number of cycles to failure ($N_f$). This adjustment is represented in Fig. \ref{fig:SNpolished}. Then, the phase field length scale $\ell$ is calculated applying Eq. \eqref{eq:AT1_ell} for the AT1 model. Notably, $\sigma_\mathrm{c}$ exceeds $\sigma_\mathrm{ULT}$. Furthermore, the endurance limit $\sigma_e$ indicated in Table \ref{table:mechprop} is also taken from Fig. \ref{fig:SNpolished}. Hence, the endurance limit $\sigma_e$ is defined as the fatigue strength at $10^6$ cycles, as in the experimental work \cite{Singh2019}. The parameter $\bar{\alpha}_{\mathrm{T}}$ is calculated using Eq. \eqref{eq:a0def}.

\subsection{Comparison with experiments}\label{sec3a}

\begin{figure}[H]
    \centering
    \includegraphics[scale=0.7]{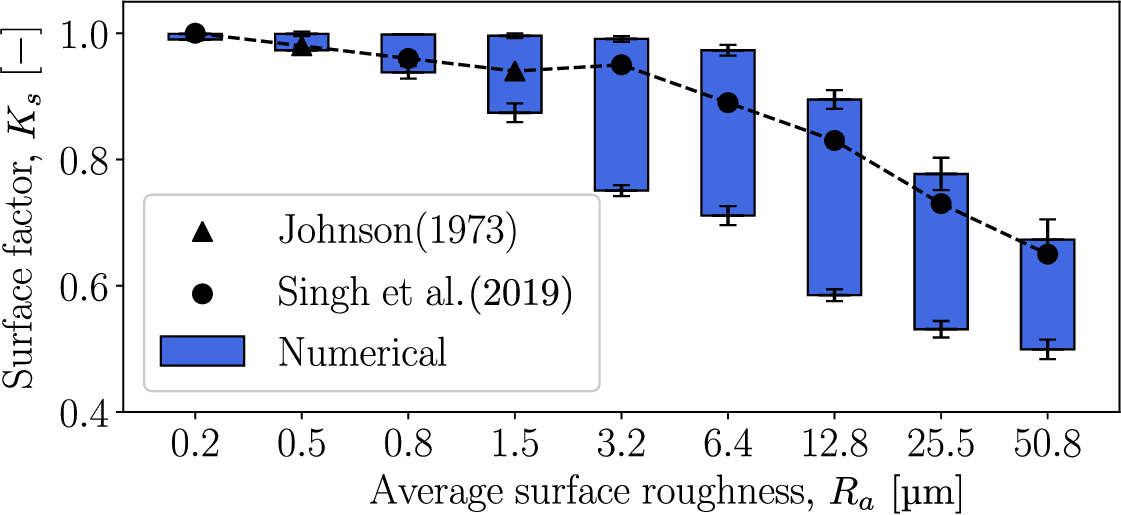}
    \caption{Comparison of the calculated surface factor to empirical results found in the literature \cite{Singh2019, Johnson1973}.}\label{fig:experiments_bar}
\end{figure}

As stated in Section \ref{sec1}, the surface factor is the primary parameter used to quantify the effect of surface roughness on the fatigue life of materials. It is defined as
\begin{equation}
    K_s = \frac{\sigma_e^\mathrm{r}}{\sigma_e^\mathrm{p}}, \label{eq:surfacefactor}
\end{equation}
where $\sigma_e^\mathrm{r}$ and $\sigma_e^\mathrm{p}$ represent the endurance limit of the rough and polished surfaces (the latter being listed in Table \ref{table:mechprop}). These variables are calculated following the procedure outlined in \ref{app1}. Fig. \ref{fig:experiments_bar} compares the values of $K_s$ obtained for AISI 4130 across different $R_a$ values. For each $R_a$, a range of possible values for $K_s$ is shown (blue range in Fig. \ref{fig:experiments_bar}), accounting for a correlation length ($\ell_\mathrm{cor}$) between $10$ and \SI{100}{\micro\meter} (typical values \cite{Bigerelle2003, Roy2018}), as well as the inherent randomness of the phenomenon. Notably, the size of these ranges generally increases for larger values of the average surface roughness $R_a$, since the effect of the correlation length becomes more significant (as assessed in Section \ref{sec3b}) and the stochastic nature of surface roughness becomes more pronounced. The figure also shows the experimental results reported by Singh et al. \cite{Singh2019} for $R_a = 0.5$ and \SI{1.5}{\micro\meter}, as well as the empirical estimates of Johnson \cite{Johnson1973} for the remaining $R_a$ values. It is worth noting that the experimental data consistently fall within the estimated range. Furthermore, as $R_a$ increases, the empirical estimates tend to align with the upper bound of the range, which corresponds to larger correlation length values.

\begin{figure} [H]
     \centering
     \begin{subfigure}[t]{0.45\textwidth}
         \centering
         \includegraphics[width=\textwidth]{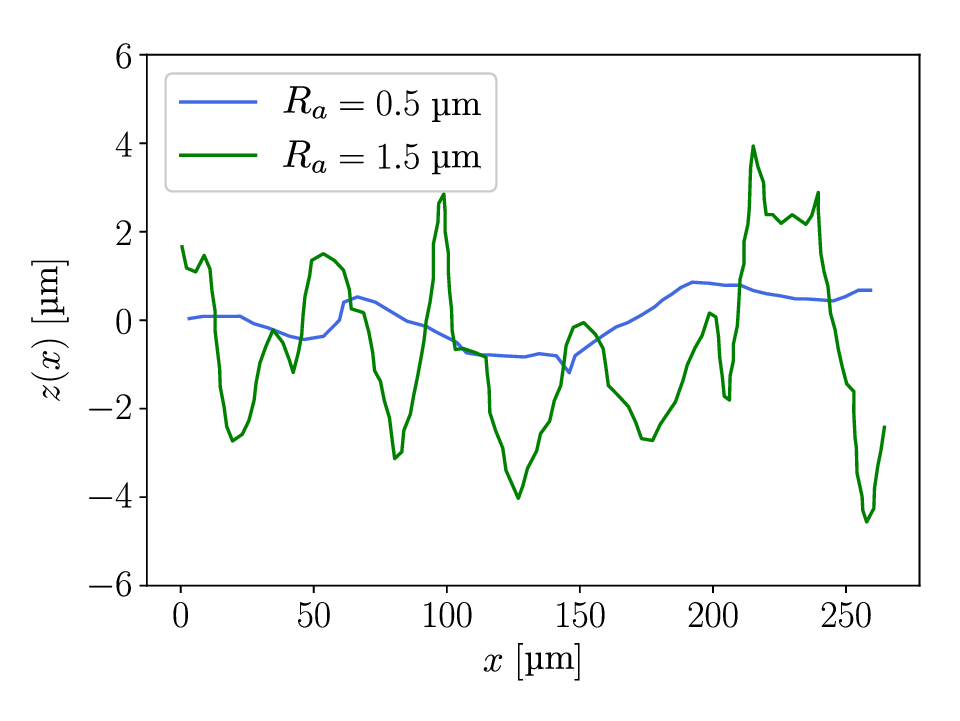}
         \caption{Surface profile}
         \label{fig:profile_exp}
     \end{subfigure}
     \hfill
	\begin{subfigure}[t]{0.45\textwidth}
         \centering
         \includegraphics[width=\textwidth]{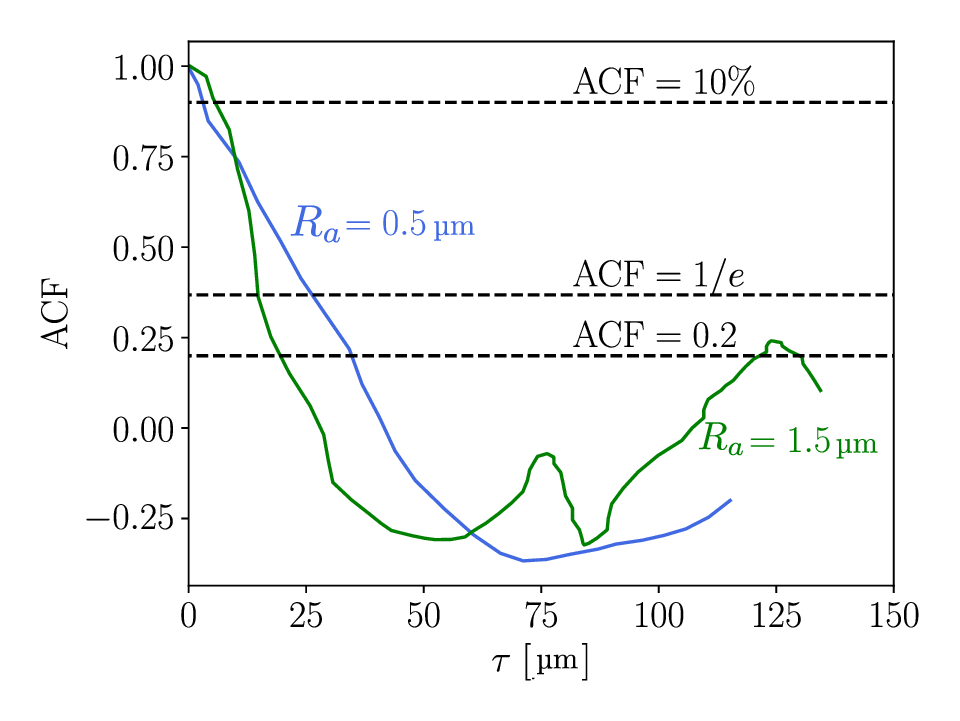}
         \caption{ACF}
         \label{fig:ACF_exp}
    \end{subfigure}
    \caption{Examples of the surface topology of the two kind of specimens tested in the experiments from Ref. \cite{Singh2019}, with $R_a=0.5$ and \SI{1.5}{\micro\meter}; (a) represents the height deviation $z(x)$ of the surface profile and (b) shows the autocorrelation functions (ACF) obtained for the two profiles, together with the three criteria used to calculate the correlation length $\ell_\mathrm{cor}$. }
        \label{fig:topology_exp}
\end{figure}

The model's accuracy improves significantly when the correlation length is known, as in the case of the experimental results from \cite{Singh2019}, where the roughness profile is available, see Fig. \ref{fig:profile_exp}. Based on these measurements an initial estimate of the ACF can be made, see Fig. \ref{fig:ACF_exp}. However, this is only a first approximation, since according to standard ISO/FDIS 2190-3, the length of the measurement domain to correctly capture the profile for $R_a=0.5$ and \SI{1.5}{\micro\meter} is larger ($d_x$ is about \SI{500}{\micro\meter}, as discussed in Section \ref{sec2b}).

Table \ref{table:exps} presents the estimated correlation length values based on the three criteria outlined in Section \ref{sec2b}, along with the corresponding experimental $K_s^{\text{exp}}$ and numerical $K_s^{\text{num}}$ estimates of the surface factor. The results indicate that the best approximation is obtained using the $\mathrm{ACF} = 0.2$ criterion, yielding a relative error of less than $1\%$. Notably, this criterion aligns with the ISO 25178 standard \cite{Blateyron2013}.

\renewcommand{\arraystretch}{1.5}
\begin{table}[h]
    \centering
    \begin{tabular}{|c|c|c|c|c|c|c|c|c|}
        \hline
        $R_a [\si{\micro\meter}]$ & \multicolumn{4}{c|}{0.5} & \multicolumn{4}{c|}{1.5} \\
        \cline{2-9}
        & $\ell_{\text{cor}}$ & $K_s^{\text{num}}$ & $K_s^{\text{exp}}$ & Error & $\ell_{\text{cor}}$ & $K_s^{\text{num}}$ & $K_s^{\text{exp}}$  & Error \\
        \hline
        ACF = 10\% & $3$ \si{\micro\meter} & $<0.97$ & & $>1\%$ & $5.5$ \si{\micro\meter} & $<0.87$ &   & 7.4\%\\
        \cline{1-3}\cline{5-7}\cline{9-0}
        ACF = 1/e & $26.4$ \si{\micro\meter} & $0.99$ & $0.98$ & $1\%$ & $14.6$ \si{\micro\meter} & $0.9 - 0.92$ & $0.94$ & 2-4\%\\
        \cline{1-3}\cline{5-7}\cline{9-0}
        ACF = 0.2 & $35.3$ \si{\micro\meter} & $0.99$ & & $1\%$ & $19.8$ \si{\micro\meter} & $0.93 - 0.95$ & & 1\%\\
        \hline
    \end{tabular}
    \caption{Surface factor obtained for the surface profiles defined in Ref. \cite{Singh2019} for $R_a=0.5$ and \SI{1.5}{\micro\meter}.}\label{table:exps}
\end{table}

\subsection{The surface factor as a function of key parameters}\label{sec3b}

One of the open questions in the literature is to quantify the surface factor ($K_s$) as a function of several key parameters: the surface topology and the tensile strength. The numerical model proposed in this paper provides the opportunity to generate novel operation maps. To achieve this, the problem outlined in Fig. \ref{fig:modelo} is considered, along with the range of values for $R_a$, $\ell_\mathrm{cor}$, and $\sigma_\mathrm{c}$. The range selected for $R_a$ is $0.2, 0.5, 0.8, 1.5, 3.2, 6.4, 12.8, 18, 25.5, 35$ and \SI{50.8}{\micro\meter}. It is based on the one studied by Johnson \cite{Johnson1973}, with some modifications. First, values of $R_a < $ \SI{0.2}{\micro\meter} are excluded, as the effects of roughness are considered negligible for these values according to ASTM standard \cite{ASTM1, ASTM2}. Second, two of these values are adjusted to include $R_a = 0.5$ and \SI{1.5}{\micro\meter}, the roughness values studied by Singh et al. \cite{Singh2019}. Finally, additional values of $R_a = 8$, $18$ and \SI{35}{\micro\meter} are included to better capture the evolution of $K_s$ with respect to $R_a$. On the other hand, the range for $\ell_\mathrm{cor}$ is based on other proposals in the literature \cite{Bigerelle2003, Roy2018}, $\ell_\mathrm{cor}=10, 20, 30, 40, 50, 60, 70, 80, 90,$ and \SI{100}{\micro\meter}. For the tensile strength $\sigma_\mathrm{c}$ the limits are inspired in Ref. \cite{Johnson1973}, $\sigma_\mathrm{c}=1121, 1500, 2000$ and $2500$ MPa.

\begin{figure}[H]
    \centering
    \includegraphics[scale=0.8]{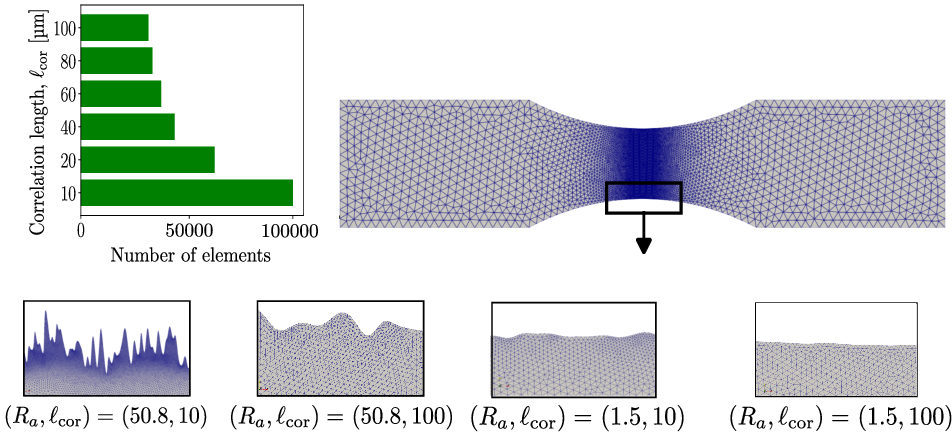}
    \caption{Mesh profiles at the border of the specimen in the critical region for different values of $R_a$ and $\ell_\mathrm{cor}$, and the number of elements in the mesh for several values of $\ell_\mathrm{cor}$ for $R_a =$ \SI{12.8}{\micro\meter}.}\label{fig:meshprofiles}
\end{figure}

By modifying the key parameters indicated above, the meshing of the numerical model is also affected, particularly in the edges of the specimen, where surface roughness is located and therefore crack nucleation is expected \cite{Singh2019}. As a general rule, the minimum element size defined at the specimen boundary $\ell_\mathrm{mesh}$ must satisfy $5\ell_\mathrm{mesh} = \min\left(  \ell_\mathrm{cor}, \ell \right)$, based on the limits established for the stochastic roughness model \cite{Loth2023} and the phase field model for fracture \cite{Kristensen2021}. Thus, $\ell_\mathrm{mesh}$ is changed when modifying $\ell_\mathrm{cor}$ and $\sigma_c$, the latter being related to $\ell$ through Eq. \eqref{eq:AT1_ell}. As an example, Fig. \ref{fig:meshprofiles} illustrates how the mesh and the number of elements change when it is controlled by $\ell_\mathrm{cor}$. While the focus of this work is on 2D plane strain problems, the framework can be readily extended to 3D by applying the method described in Section \ref{sec2b} across the entire boundary surface.

\begin{figure}[H]
    \centering
    \includegraphics[scale=0.8]{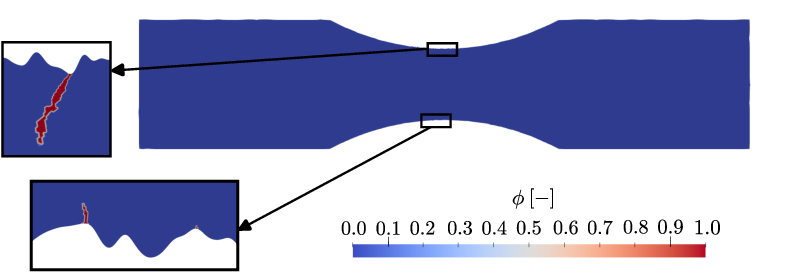}
    \caption{Nucleation of cracks at the stress concentrators captured by the model. This is shown for $R_a=$ \SI{12.8}{\micro\meter}, $l_\mathrm{cor}=$ \SI{50}{\micro\meter} and $\sigma_\mathrm{c}=2500$ MPa.}\label{fig:damage}
\end{figure}

In the numerical simulation, the failure of the specimen begins at the stress concentrators generated by the surface roughness, where small cracks nucleate and propagate into the interior of the specimen, as shown in Fig. \ref{fig:damage}. This analysis is a force-controlled test, and therefore the number of cycles to failure is determined by the cycle in which catastrophic failure occurs, characterised by a sudden attainment of $\phi=1$ in a large fraction of the specimen.  

\begin{figure} [H]
     \centering
     \begin{subfigure}[t]{0.42\textwidth}
         \centering
         \includegraphics[width=\textwidth]{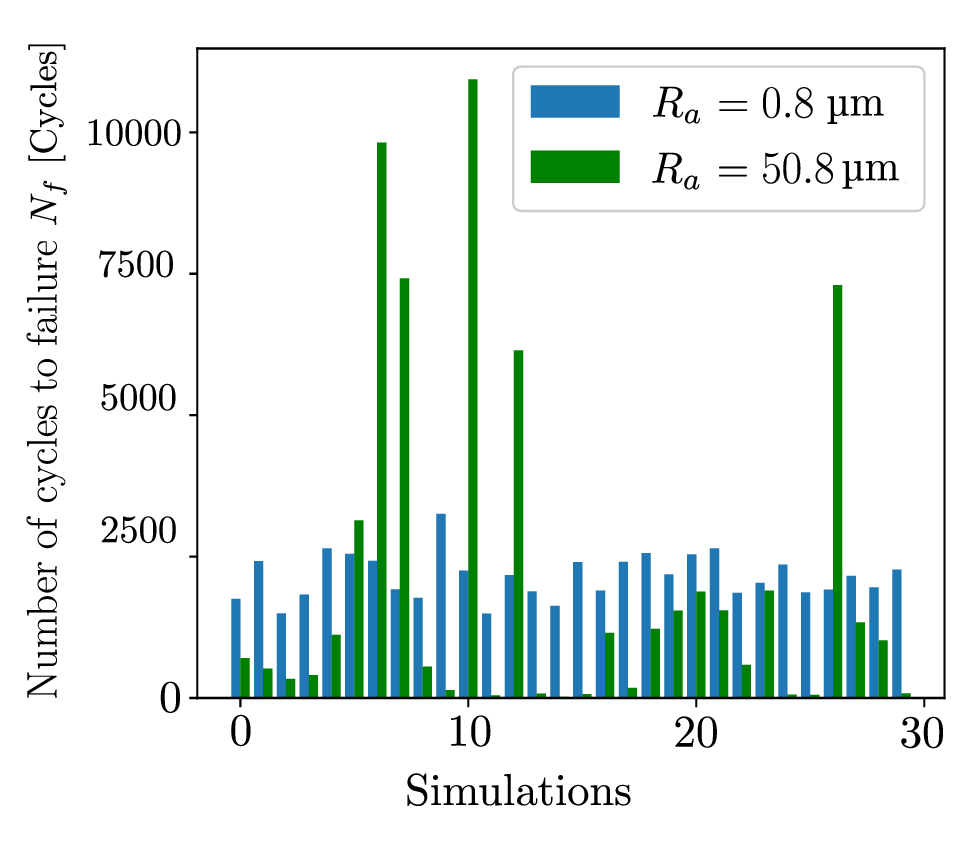}
         \caption{}
         \label{fig:sample}
     \end{subfigure}
     \hfill
	\begin{subfigure}[t]{0.45\textwidth}
         \centering
         \includegraphics[width=\textwidth]{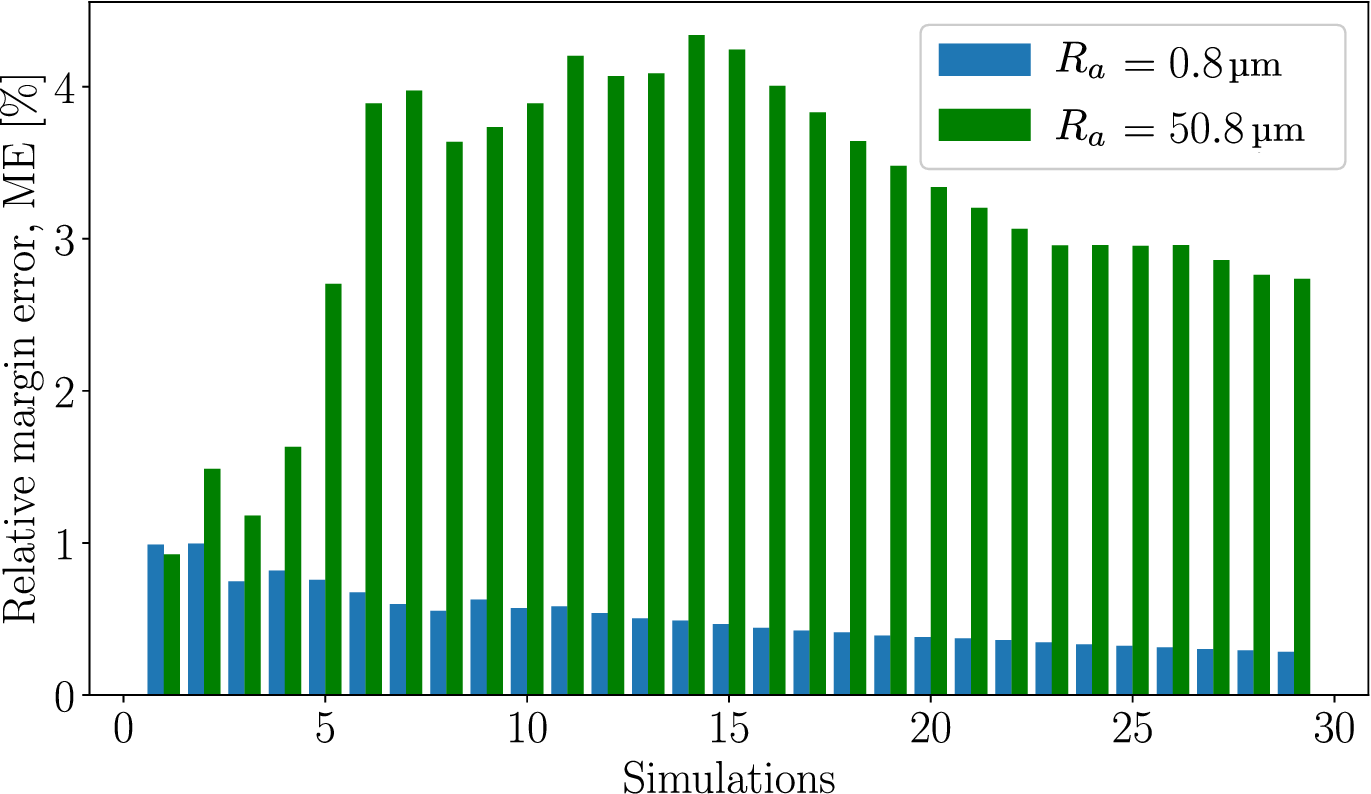}
         \caption{}
         \label{fig:sample_me}
    \end{subfigure}
    \caption{Evolution of $N_f$ and $\mathrm{ME}$ in a sample for $R_a=0.8$ and \SI{50.8}{\micro\meter}, $\ell_\mathrm{cor}=$ \SI{20}{\micro\meter} and $\sigma_\mathrm{c}=1121$ MPa.}
        \label{fig:sampleandme}
\end{figure}

Due to the stochastic nature of this phenomenon, a single simulation per case is insufficient to obtain a reliable result, as illustrated in Fig. \ref{fig:sample}, which shows the number of cycles to failure $N_f$ for two different roughness levels $R_a = 0.8$ and $R_a =$ \SI{50.8}{\micro\meter}. To compare both cases in the same graph, a stress amplitude of $\sigma_a=340$ and $\sigma_a=230$ MPa is applied for the $R_a = 0.8$ and $R_a =$ \SI{50.8}{\micro\meter} cases, respectively, ensuring that the average number of cycles to failure is approximately $3000$ cycles. This allows highlighting the differences in variability induced solely by the roughness parameter. In both cases, we assume $\ell_\mathrm{cor}=$ \SI{20}{\micro\meter} and $\sigma_\mathrm{c}=1121$ MPa. In this context, to ensure the reliability of the results, it is worth measuring the margin of error of the sample, see Eq. \eqref{eq:me}, which decreases as the sample size increases, as observed in Fig. \ref{fig:sample_me}. To achieve a Gaussian distribution in the results, at least 30 simulations per case are performed (as indicated in Section \ref{sec2b}), ensuring a margin of error below $5 \%$. As shown in Fig. \ref{fig:sample_me}, this requirement is met in all cases with 30 simulations. 

\begin{figure} [H]
     \centering
     \begin{subfigure}[t]{0.45\textwidth}
         \centering
         \includegraphics[width=\textwidth]{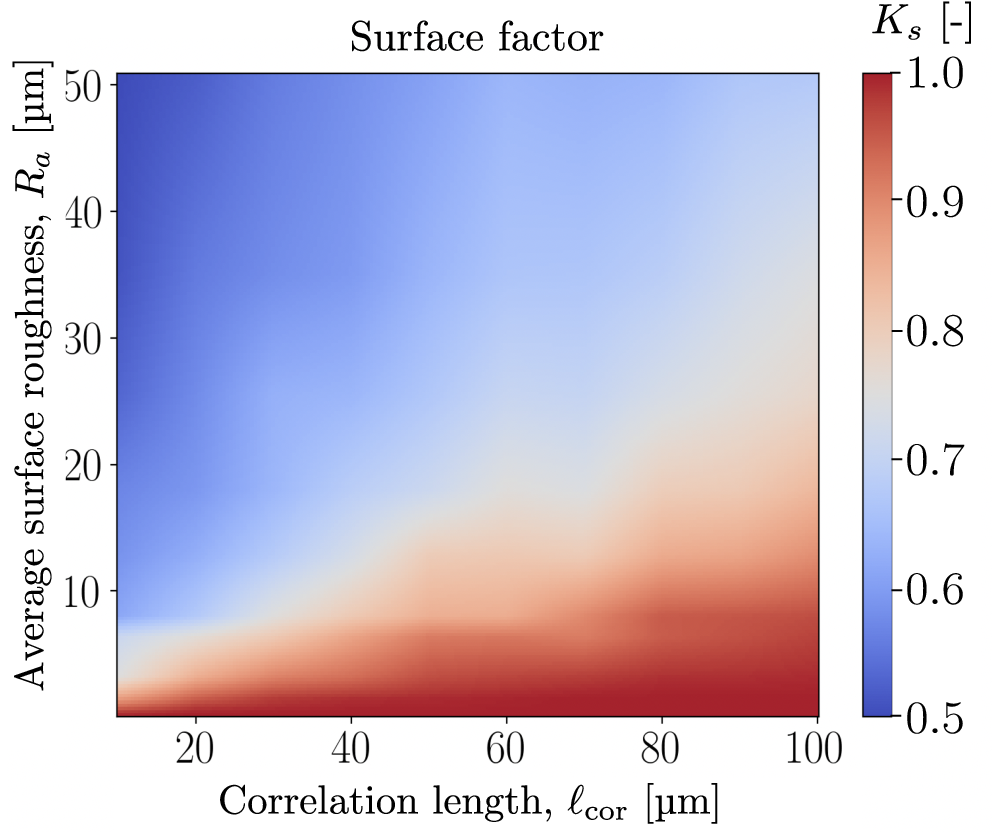}
         \caption{}
         \label{fig:mapRL_K}
     \end{subfigure}
     \hfill
	\begin{subfigure}[t]{0.48\textwidth}
         \centering
         \includegraphics[width=\textwidth]{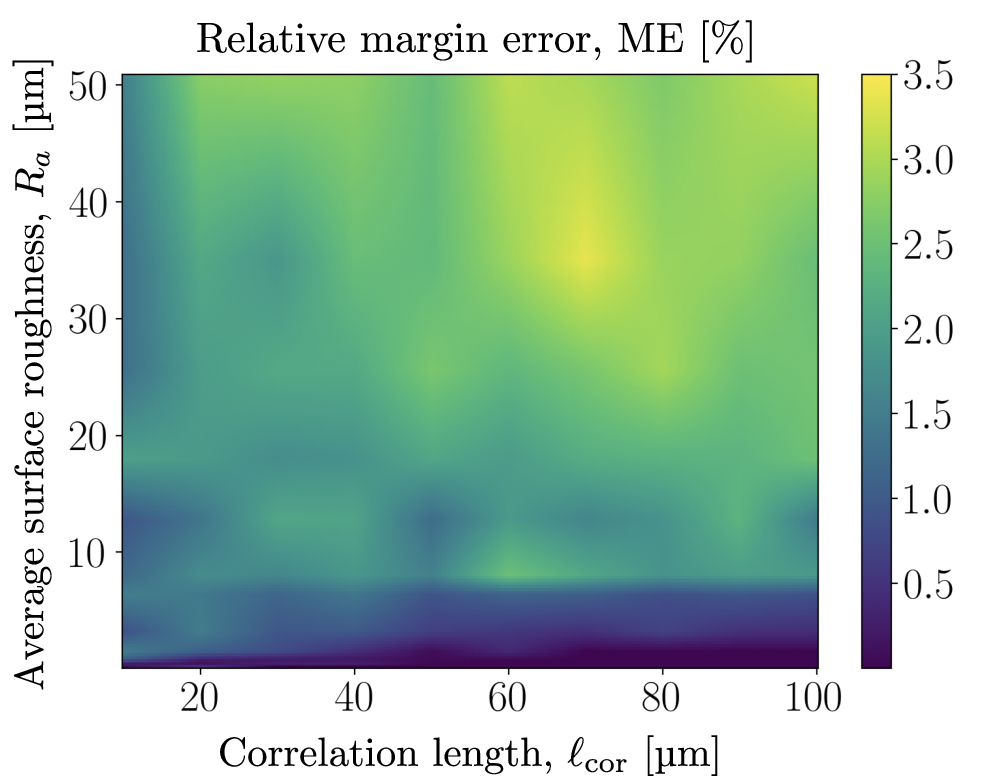}
         \caption{}
         \label{fig:mapRL_ME}
    \end{subfigure}
    \caption{Influence of the surface roughness and the correlation length on the surface factor $K_s = \frac{\sigma_e^\mathrm{r}}{\sigma_e^\mathrm{p}}$ for $\sigma_\mathrm{c}=1121$ MPa.}
        \label{fig:mapRL}
\end{figure}

Fig. \ref{fig:mapRL}, a key result of this work, shows the influence of surface topology on the surface factor. This map has been built by running 110 case studies (with 30 simulations per case study) corresponding to the range of $\ell_\mathrm{cor}$ and $R_a$ discussed at the beginning of this section. In Fig. \ref{fig:mapRL_K}, the evolution of $K_s$ is represented as a function of $\ell_\mathrm{cor}$ and $R_a$. The biggest values (closer to $1$), which indicate a lower influence of roughness on the fatigue life of the specimen, correspond to larger $\ell_\mathrm{cor}$ and smaller $R_a$ values. Conversely, the smallest $K_s$ are obtained for smaller $\ell_\mathrm{cor}$ and larger $R_a$ values. The smallest $K_s$ values attained reach approximately $0.5$, which aligns with the minimum values empirically estimated in the literature \cite{Johnson1973,McKelvey2012}. On the other hand, Fig. \ref{fig:mapRL_ME} shows the evolution of the sample's margin of error. It is observed that for large $\ell_\mathrm{cor}$ and large $R_a$, the margin of error is higher, while it is nearly negligible for small average roughness values. This implies that as roughness increases, the stochastic characteristics of the problem become more critical, making it easier to obtain varying results. Consequently, considering a larger sample size becomes essential. Interestingly, Fig. \ref{fig:mapRL_ME} shows a clear increase in the margin of error for $R_a > 8\ \si{\micro\meter}$, particularly at larger correlation lengths. For $R_a \leq 8\ \si{\micro\meter}$, the influence of roughness on fatigue is smaller, as the coefficient $K_s$ remains between 0.8 and 1 (see Fig. \ref{fig:mapRL_K}), resulting in a very small margin of error. This shift highlights the growing impact of surface features at higher roughness levels.

\begin{figure}[H]
    \centering
    \includegraphics[scale=0.55]{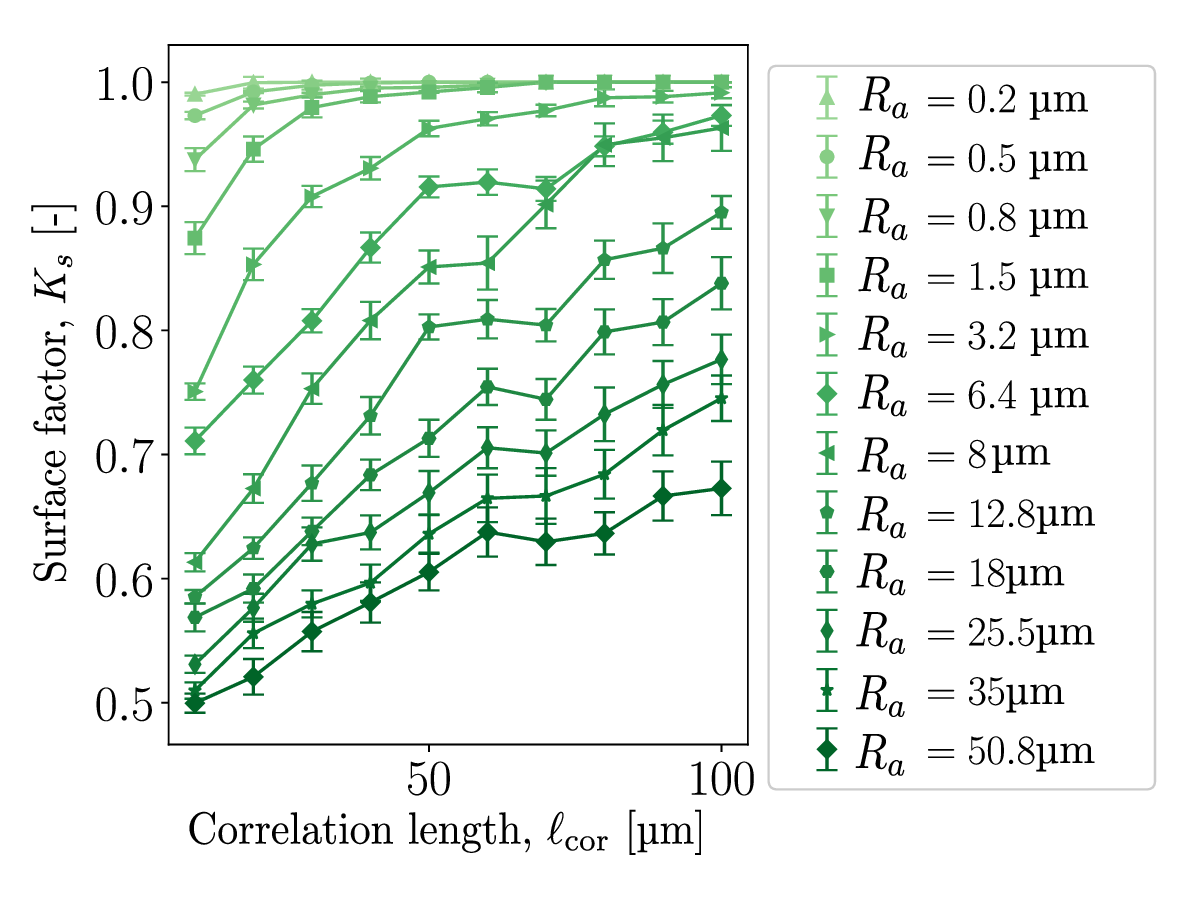}
    \caption{Evolution of the surface factor with respect to the correlation length for several values of the average surface roughness for $\sigma_\mathrm{c}=1121$ MPa.}\label{fig:mapKalcorRab}
\end{figure}

The stochastic nature of the roughness phenomenon can also be observed in Fig. \ref{fig:mapKalcorRab}, where $K_s$ is represented as a function of the correlation length for the selected range of the average surface roughness. It is evident that $K_s$ does not follow a smooth evolution with respect to $\ell_\mathrm{cor}$ and $R_a$ due to the presence of the margin of error, which is depicted as a range of possible values for each studied point. For small roughness values, it is observed that beyond a certain $\ell_\mathrm{cor}$, the effect of roughness becomes negligible.

\begin{figure}[H]
    \centering
    \includegraphics[scale=0.55]{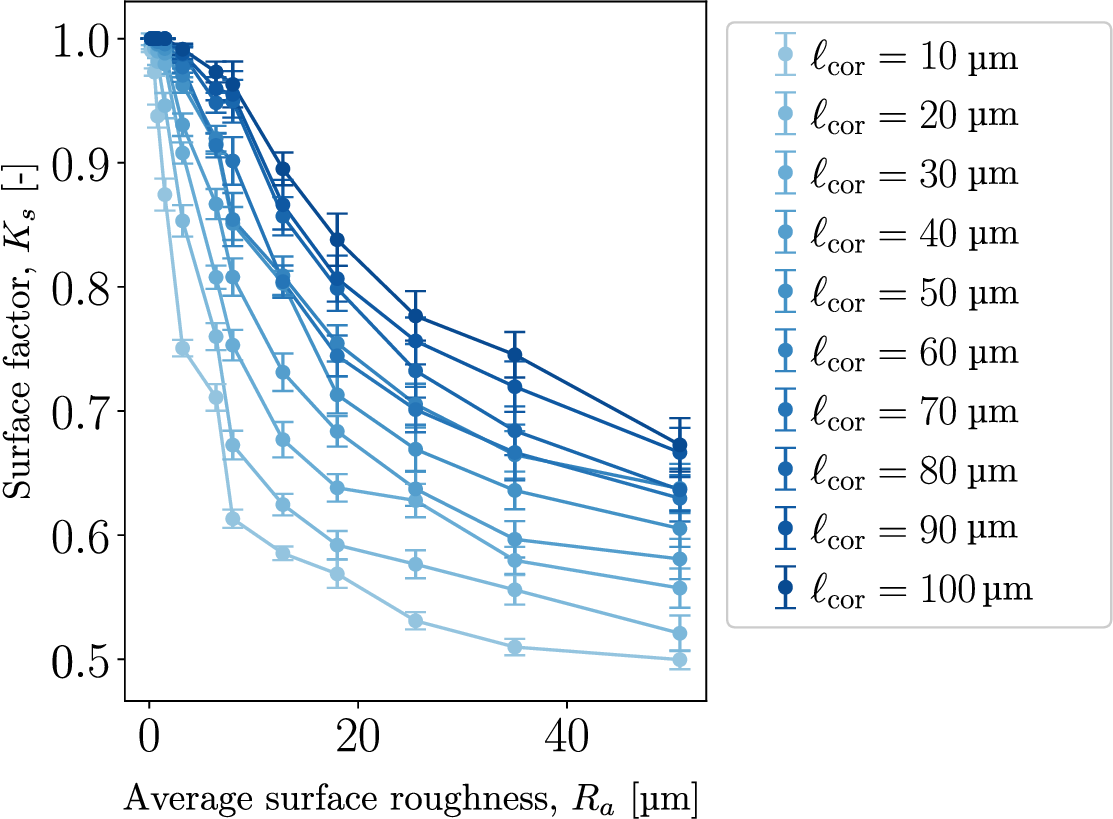}
    \caption{Evolution of the surface factor with respect to the average surface roughness for several values of the correlation length for $\sigma_\mathrm{c}=1121$ MPa.}\label{fig:mapKalcorRabII_natural}
\end{figure}

The order of magnitude of $R_a$ seems to play a significant role in the criticality of surface roughness, as shown in Fig. \ref{fig:mapKalcorRabII_natural}, where $K_s$ is represented as a function of $R_a$ for several values of $\ell_\text{cor}$. Larger $R_a$ leads to lower values of $K_s$. In fact, for $R_a$ on the order of unity, the surface factor exhibits small variations and remains close to $1$. However, for $R_a$ values on the order of ten, the effect of $R_a$ becomes more critical, reducing the surface factor to minimum values of 0.5. Notably, $K_s = 0.5$ implies that the fatigue strength at $10^6$ cycles of the rough component is reduced by half compared to the same polished specimen, representing a significant change in design conditions.

In the literature, several authors have proposed estimations of $K_s$ as a function of the manufacturing process \cite{Shigleys2011} or material properties, particularly the ultimate tensile strength \cite{Johnson1973, McKelvey2012}. In this paper, the influence of fracture strength is studied using the values proposed above, $\sigma_\mathrm{c}=1121, 1500, 2000$ and $2500$ MPa. First, it is necessary to modify the fatigue test used to estimate the endurance limit of the polished surface for each of these values, as the change in one of the material properties affects the estimation of the endurance limit $\sigma_\mathrm{e}^\mathrm{p}$. Assuming that modifying $\sigma_\mathrm{c}$ does not change the slope of the S-N curve, the parameter $a$ of the Basquin's law is adjusted through a small benchmark test: by applying a certain stress $\sigma_i = a_i N_{fi}^{-b}$, the expected number of cycles to failure $N_f$ should match exactly the one used to define the imposed load, $N_f = N_{fi}$. Table \ref{table:polished_stress} presents the obtained values of the phase field length scale $\ell$, the coefficient $a$, the coefficient $\alpha_{\mathrm{T}}$, and the endurance limit $\sigma_\mathrm{e}$ for each $\sigma_\mathrm{c}$ in the range. These values are also used in the test of the rough specimen. Interestingly, the parameter $\alpha_{\mathrm{T}}$ is the same for $\sigma_\text{c}=2000$ and $\sigma_\text{c}=2500$ MPa. This is uncommon but can occur due to the small variation in the coefficient $a$ compared to the other cases. This may reduce the accuracy of the model, as it assumes that a change in the fracture strength does not modify certain parameters (like $b$), and the model is not fitted with any real polished S-N curve.

\begin{figure} [H]
     \centering
     \begin{subfigure}[t]{0.45\textwidth}
         \centering
         \includegraphics[width=\textwidth]{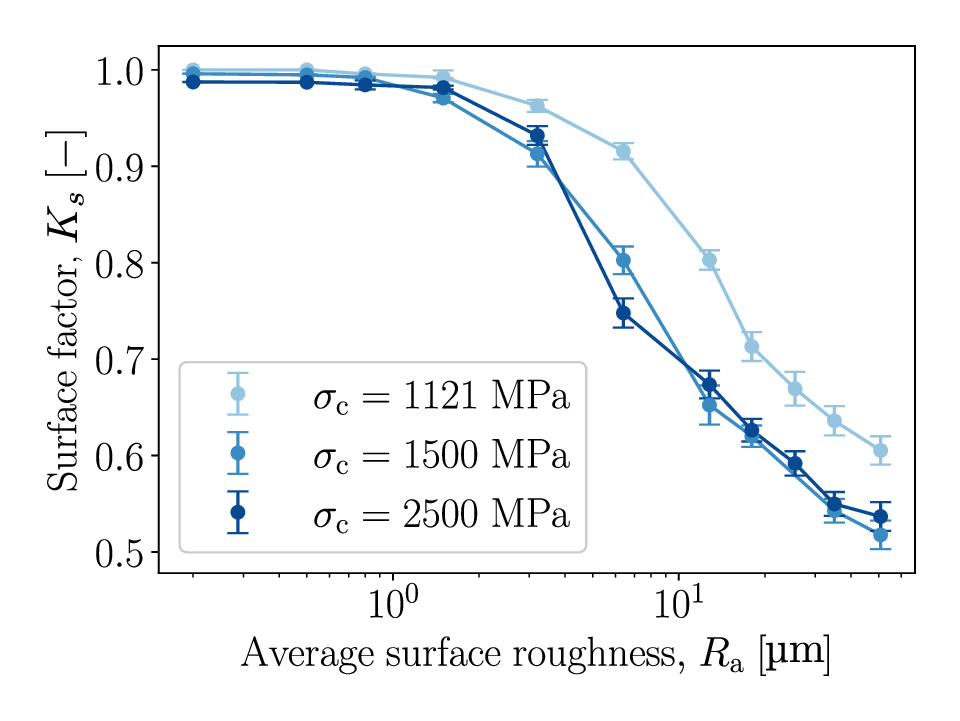}
         \caption{Evolution of the surface factor with respect to the average surface roughness for several values of $\sigma_\mathrm{c}$. Notice that $R_a$ is represented on a logarithmic scale.}
         \label{fig:mapKRaSc}
     \end{subfigure}
     \hfill
	\begin{subfigure}[t]{0.51\textwidth}
         \centering
         \includegraphics[width=\textwidth]{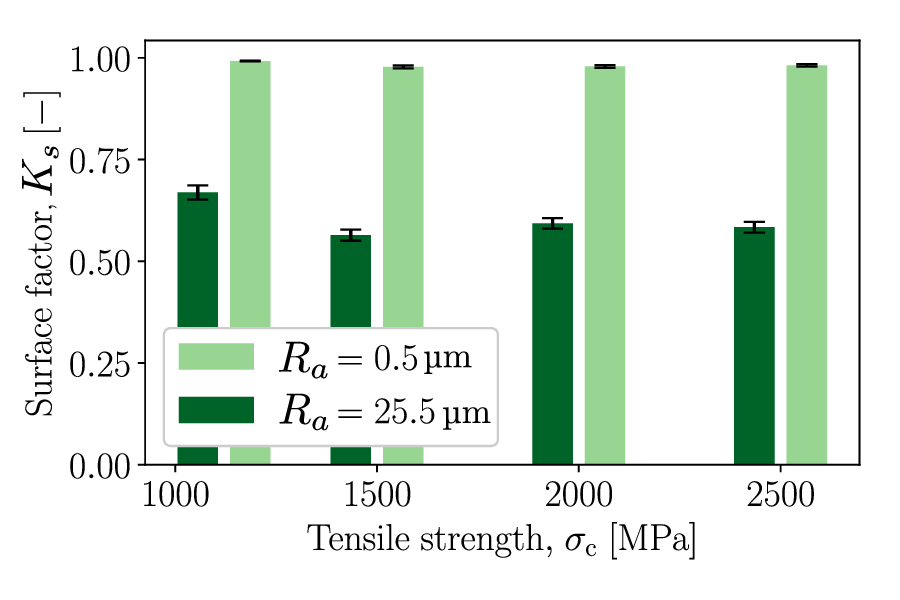}
         \caption{Evolution of the surface factor with respect to the average surface roughness for several values of $\sigma_\mathrm{c}$ and $R_a$. }
         \label{fig:mapKRaScII}
    \end{subfigure}
    \caption{Influence of the fracture strength in the surface factor. The correlation length in the analysis is $\ell_\mathrm{cor} = $ \SI{50}{\micro\meter}. }
        \label{fig:mapSc}
\end{figure}

Fig. \ref{fig:mapKRaSc} shows the evolution of the surface factor with $R_a$ for three different values of $\sigma_\mathrm{c}$. For $R_a$ values on the order of unity, a very small dependence of $K_s$ on $\sigma_\mathrm{c}$ is observed. However, for high roughness values, a change in $\sigma_\mathrm{c}$ can significantly reduce the surface factor. The criticality of this parameter becomes negligible for high $\sigma_\mathrm{c}$ values. In fact, between $\sigma_\mathrm{c}=1500$ MPa and $\sigma_\mathrm{c}=2500$  MPa, the effect is so small that several points overlap due to the randomness of the phenomenon. These conclusions were also observed in Ref. \cite{Johnson1973} with respect to $\sigma_\mathrm{ULT}$, leading to the conclusion that the largest variations occur at high roughness and low strength levels.

\begin{table}[H]
    \centering
    \begin{tabular}{|c|c|c|c|c|}
        \hline
        $ \sigma_c $ [MPa] & $1121$ & $1500$ & $2000$ & $2500$ \\
        \hline
        $ a $ [MPa] & $485.9$ & $960.22$ & $1660$ & $2075$ \\
        \hline
        $ \sigma_e $ [MPa] & $263$ & $521$ & $902$ & $1126$ \\
        \hline
        $\bar{\alpha}_\text{T}$ & $8.2 \times 10^{-4}$ & $7.47$ & $4234$ & $4234$ \\
        \hline
        $ \ell $ [mm] & $2.9$ & $1.62$ & $0.91$ & $0.58$ \\
        \hline
    \end{tabular}
    \caption{Table of material parameters for different values of $ \sigma_\mathrm{c} $.}
    \label{table:polished_stress}
\end{table}

The effect of fracture strength is shown for two very different roughness values in Fig. \ref{fig:mapKRaScII}, highlighting the error margins. In general, no significant changes in randomness are observed as $\sigma_\mathrm{c}$ increases. On the other hand, as mentioned earlier, between $1000$ and $1500$ MPa, a more noticeable reduction is observed; however, beyond $1500$ MPa, the changes become negligible and are influenced by the effect of randomness.

The tool presented in this study estimates the effect of surface roughness on high-cycle fatigue. The accuracy of these estimations improves with additional data on material properties and roughness characteristics, including correlation length and average roughness. However, when certain parameters, such as correlation length, are unavailable, reasonable approximations can be made. As discussed in Section \ref{sec1}, additively manufactured materials are particularly sensitive to surface roughness, which is strongly influenced by the manufacturing process. For instance, in Laser Powder Bed Fusion (LPBF), typical $R_a$ values range from 18 to \SI{21}{\micro\meter} \cite{Shrestha2019}. According to the map in Fig. \ref{fig:mapRL_K}, for an ultimate strength of 530 MPa, the surface factor for $R_a$ between 18 and \SI{21}{\micro\meter} falls within the range of 0.56 to 0.84, with an average value of 0.7. Notice that the same range for the correlation length has been chosen, based on the values reported in \cite{Thompson2018, Snyder2020}. This estimate closely aligns with the experimental findings reported in Ref. \cite{Shrestha2019}, which examines the fatigue behavior of additively manufactured 316L stainless steel manufactured by LPBF with an ultimate strength of 587 MPa. Experimental data indicate that for an as-built surface with $R_a$ between 18 and \SI{21}{\micro\meter}, failure at $10^6$ cycles occurs at an applied stress of $\sigma=133$ MPa, corresponding to a surface roughness factor of 0.77—less than $10\%$ deviation from the averaged value predicted by our tool. Another application where the surface topology becomes very critical is in the fatigue life of specimens and components that have been subjected to pitting corrosion. In Ref. \cite{PerezMora2015}, the effect of pitting corrosion on the fatigue life of R5 steel specimens was experimentally investigated, considering pits with an average size of \SI{50}{\micro\meter}. In comparison to non-corroded specimens, the fatigue strength at which specimens failed at $10^7$ cycles decreased from 390 to 300 MPa. In Ref. \cite{Chauhan2023} a typical value of $\ell_\mathrm{cor} = 300-400$ \si{\micro\meter} is reported for $R_a \approx 50$ \si{\micro\meter}, obtaining an average reduction from 390 to 348 MPa under those conditions; i.e., a difference of $16\%$. Notice that the difference with experimental values is higher in this example because R5 has an ultimate strength of 1018 MPa, and the estimation is provided through the map in Fig. \ref{fig:mapRL_K} for a steel of $\sigma_\mathrm{ULT}=530$ MPa.

\section{Conclusions}\label{sec4}

This paper presents a new computational framework for predicting the fatigue life of rough components in the high cycle fatigue (HCF) regime. Its main advantage is that it requires only basic parameters of the roughness profile while accounting for the stochastic nature of the phenomenon under study. Among other capabilities, it allows for the estimation of the surface factor, the most widely used parameter to quantify the influence of roughness on fatigue.  Key findings include:

\begin{itemize}

    \item Model predictions are in very good qualitative and quantitative agreement with the experimental literature. The accuracy is very high when the roughness profile is known (i.e., a correlation length and an average surface roughness). According to the results of this study, the $0.2$ criterion for defining the correlation length provides the best agreement with experimental results. When the correlation length is unknown, the model is capable of providing a range of values that approximate the experimental findings.  

    \item A surface vs fatigue life operational map has been established, showing a very good agreement with experiments. In this paper, the tool has been applied to steel, as it is the most extensively studied metal in terms of surface roughness and has the most available data, allowing for a thorough validation of the computational framework. However, the model can also be applied to other metals, such as aluminum, revealing the behaviour of new materials under rough surface conditions, making it a valuable design tool. 

    \item The analysis shows that a small correlation length and a high average surface roughness increase the severity of the phenomenon, potentially halving the fatigue strength of the component. On the other hand, increasing the fracture strength reduces the surface factor, although this effect becomes negligible as the tensile strength rises above a certain level. 
    
    \item The study effectively incorporates the stochastic nature of surface roughness, demonstrating how randomness affects sample results and how to obtain reliable outcomes while incorporating this inherent characteristic of rough surfaces. To account for this, the tool generates rough profiles following a Gaussian distribution. Although the current approach relies on explicit modelling of roughness, future implementations could benefit from incorporating statistical frameworks—such as weakest link models.

\end{itemize}

The computational approach presented offers a robust methodology for estimating the surface factor across various applications where surface roughness plays a particularly critical role. In this study, specific estimates are provided for the reduction in fatigue strength at one million cycles in the contexts of additive manufacturing and pitting corrosion. These estimations align well with established experimental data from the literature. Moreover, the tool could be applied to generate an accurate numerical database for a machine learning model to predict the effect of surface roughness in metals, reducing the need for costly experimental data. \\

\noindent\textbf{Acknowledgments}

S. Jimenez-Alfaro acknowledges the Iberdrola Foundation under the Marie Skłodowska-Curie Grant Agreement No 101034297. E. Martínez-Pañeda acknowledges financial support from the EPSRC Supergen ORE Hub (Grant FF2023-1028) and from UKRI’s Future Leaders Fellowship programme [grant MR/V024124/1].

\appendix
\section{Further details of numerical analysis} \label{app1}

One of the key aspects of this study is the determination of the endurance limit (fatigue strength at one million cycles, according to the material properties for AISI 4130 reported in Ref. \cite{Singh2019}). One of the main challenges in fatigue analysis is its high computational cost, as indicated in Table \ref{table:computacion}. This cost increases when considering rough specimens for two reasons: (i) the number of mesh elements increases, and (ii) the stochastic nature of the phenomenon requires multiple simulation runs to obtain reliable results. For instance, determining the number of cycles to failure (with an average of 37000 cycles) under an applied stress of $270$ MPa for a rough specimen with $R_a=$ \SI{1.5}{\micro\meter} and an average of 45880 elements requires approximately 25 hours of computation. If 30 specimens are needed, obtaining just a single data point would require a total simulation time of one month. 


\begin{table}[h]
    \centering
    \renewcommand{\arraystretch}{1.5} 
    \begin{tabular}{|c|c|c|c|c|}
        \hline
        \textbf{Case} & \textbf{Elements} & $\mathbf{\sigma}$ \textbf{[MPa]} & \textbf{Cycles} & \textbf{Computation time [h]} \\
        \hline
        $R_a =$ \SI{1.5}{\micro\meter} & $45880$ & $270$ & $37425$ & $25.73$ \\
        \hline
        Polished & $30524$ & $340$ & $3000$ & $38.8$ \\
        \hline
    \end{tabular}
    \caption{Table with computation times for a rough and a polished case. For the roughness case, $\sigma_\mathrm{c}=1121$ MPa and $\ell_\mathrm{cor}=$ \SI{30}{\micro\meter} are considered.}
    \label{table:computacion}
\end{table}

\begin{figure}[H]
    \centering
    \includegraphics[scale=0.6]{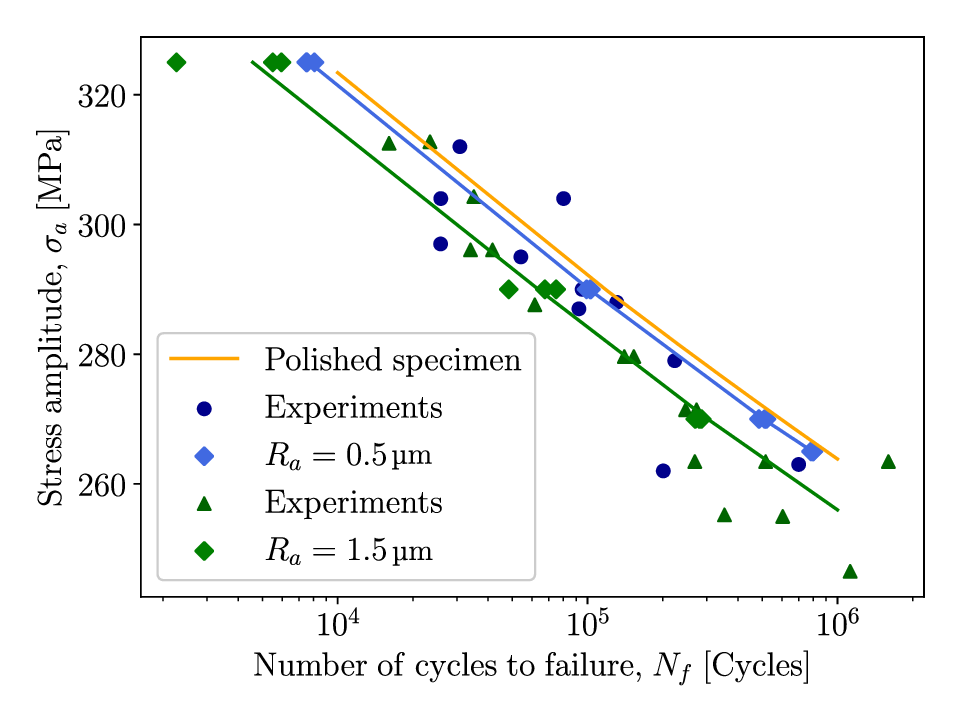}
    \caption{SN curves for $R_a = 0.5, $ \SI{1.5}{\micro\meter}. Comparison to experiments \cite{Singh2019}. }\label{fig:SNcomparison}
\end{figure}

Since this would significantly hinder the study, an alternative and reliable approach is proposed to reduce the computational cost. As shown in Fig. \ref{fig:SNcomparison}, the model reproduces parallel curves when roughness is implemented. This approach is realistic and accurate in the HCF regime, as the effect of plasticity is negligible \cite{Singh2019}. Therefore, for a given roughness level, the stress corresponding to an average of $3000$ cycles will be used for the calculation of $\sigma_e^\mathrm{r}$, applying the Basquin's law (see Section \ref{sec2}).  

\biboptions{sort&compress}


\begin{thebibliography}{10}
\expandafter\ifx\csname url\endcsname\relax
  \def\url#1{\texttt{#1}}\fi
\expandafter\ifx\csname urlprefix\endcsname\relax\def\urlprefix{URL }\fi
\expandafter\ifx\csname href\endcsname\relax
  \def\href#1#2{#2} \def\path#1{#1}\fi

\bibitem{Shigleys2011}
R.~G. Budynas, J.~K. Nisbett, J.~E. Shigley, {Shigley's mechanical engineering
  design}, McGraw-Hill, 2011.

\bibitem{Suresh1998}
S.~Suresh, {Fatigue of Materials}, Cambridge University Press, 1998.

\bibitem{Marin1962}
J.~Marin, {Mechanical behavior of engineering materials}, Prentice-Hall, 1962.

\bibitem{Maiya1975}
P.~S. Maiya, D.~E. Busch, {Effect of Surface Fatigue Behavior Roughness on
  Low-Cycle of Type 304 Stainless Steel}, Metallurgical Transactions A 6 (1975)
  1761--1766.

\bibitem{Bayoumi1995}
M.~R. Bayoumi, A.~K. Abdellatif, {Effect of surface finish on fatigue
  strength}, Engineering Fracture Mechanics 51~(5) (1995) 861--870.

\bibitem{Fordham1997}
J.~D. Fordham, R.~Pilkington, C.~C. Tang, {The effect of different profiling
  techniques on the fatigue performance of metallic membranes of AISI 301 and
  Inconel 718}, International Journal of Fatigue 19~(6) (1997) 487--501.

\bibitem{Murakami2002}
Y.~Murakami, {Effect of surface roughness on fatigue strength}, in: Y.~Murakami
  (Ed.), Metal Fatigue: Effect of Small Defects and Non Metallic Inclusions,
  Academic Press, 2019, pp. 407--430.

\bibitem{Arola2002}
D.~Arola, C.~L. Williams, {Estimating the fatigue stress concentration factor
  of machined surfaces}, International Journal of Fatigue 24 (2002) 923--930.

\bibitem{Itoga2003}
H.~Itoga, K.~Tokaji, M.~Nakajima, H.~N. Ko, {Effect of surface roughness on
  step-wise S-N characteristics in high strength steel}, International Journal
  of Fatigue 25~(5) (2003) 379--385.

\bibitem{Suraratchai2008}
M.~Suraratchai, J.~Limido, C.~Mabru, R.~Chieragatti, {Modelling the influence
  of machined surface roughness on the fatigue life of aluminium alloy},
  International Journal of Fatigue 30~(12) (2008) 2119--2126.

\bibitem{Singh2019}
K.~Singh, F.~Sadeghi, M.~Correns, T.~Blass, {A microstructure based approach to
  model effects of surface roughness on tensile fatigue}, International Journal
  of Fatigue 129 (2019) 105229.

\bibitem{Vayssette2019}
B.~Vayssette, N.~Saintier, C.~Brugger, M.~El~May, E.~Pessard, {Numerical
  modelling of surface roughness effect on the fatigue behavior of Ti-6Al-4V
  obtained by additive manufacturing}, International Journal of Fatigue 123
  (2019) 180--195.

\bibitem{Gustafsson2024}
D.~Gustafsson, S.~Parareda, R.~Munier, E.~Olsson, {High cycle fatigue life
  estimation of punched and trimmed specimens considering residual stresses and
  surface roughness}, International Journal of Fatigue 186 (2024) 108384.

\bibitem{As2005}
S.~K. {\AA}s, B.~Skallerud, B.~W. Tveiten, B.~Holme, {Fatigue life prediction
  of machined components using finite element analysis of surface topography},
  International Journal of Fatigue 27~(10-12) (2005) 1590--1596.

\bibitem{Noll1946}
C.~Noll, C.~Lipson, {Allowable working stresses}, Conference Proceedings of the
  Society for Experimental Mechanics Series 3~(2) (1946) 89--109.

\bibitem{Juvinall2020}
R.~Juvinall, K.~Marshek, {Fundamentals of machine component design}, John Wiley
  {\&} Sons, 2020.

\bibitem{Johnson1973}
R.~C. Johnson, {Specifying a surface finish that won't fail in fatigue}, in:
  Machine Design, Vol.~45, 1973, Ch.~11, pp. 108--109.

\bibitem{Rennert2024}
R.~Rennert, M.~Vormwald, A.~Esderts, {FKM-guideline “Analytical strength
  Assessment” – Background and current developments}, International Journal
  of Fatigue 182 (2024) 108165.

\bibitem{Deng2009}
G.~Deng, K.~Nagamoto, Y.~Nakano, T.~Nakanishi, {Evaluation of the effect of
  surface roughness on crack initiation life}, in: Proceedings of the 12th
  International Conference on Fracture, 2009, pp. 1--8.

\bibitem{McKelvey2012}
S.~A. McKelvey, A.~Fatemi, {Surface finish effect on fatigue behavior of forged
  steel}, International Journal of Fatigue 36~(1) (2012) 130--145.

\bibitem{Shareef1996}
I.~Shareef, M.~D. Hasselbusch, {Endurance limit modifying factors for hardened
  machine surfaces}, SAE Transactions 105 (1996) 889--899.

\bibitem{Sinclair1957}
B.~G.~M. Sinclair, H.~T. Corten, T.~J. Dolan, {Effect of Surface Finish on the
  Fatigue Strength of Titanium Alloys RC 130B and Ti 140A}, Transactions of the
  American Society of Mechanical Engineers 79~(1) (1957) 89--95.

\bibitem{Solberg2019}
K.~Solberg, S.~Guan, N.~Razavi, T.~Welo, K.~C. Chan, F.~Berto, {Fatigue of
  additively manufactured 316L stainless steel: The influence of porosity and
  surface roughness}, Fatigue and Fracture of Engineering Materials and
  Structures 42~(9) (2019) 2043--2052.

\bibitem{Lee2020}
S.~Lee, J.~W. Pegues, N.~Shamsaei, {Fatigue behavior and modeling for additive
  manufactured 304L stainless steel: The effect of surface roughness},
  International Journal of Fatigue 141 (2020) 105856.

\bibitem{Zhang2019}
J.~Zhang, A.~Fatemi, {Surface roughness effect on multiaxial fatigue behavior
  of additive manufactured metals and its modeling}, Theoretical and Applied
  Fracture Mechanics 103 (2019) 102260.

\bibitem{Neuber1961}
H.~Neuber, {Theory of Stress Concentration for Shear-Strained Prismatical
  Bodies With Arbitrary Nonlinear Stress-Strain Law}, Journal of Applied
  Mechanics 28~(4) (1961) 544--550.

\bibitem{Peterson1959}
R.~Peterson, {Notch Sensitivity}, in: G.~Sines, J.~Waisman (Eds.), Metal
  Fatigue, Mc Graw-Hill, 1959, pp. 293--306.

\bibitem{Murakami1983}
Y.~Murakami, M.~Endo, {Quantitative evaluation of fatigue strength of metals
  containing various small defects or cracks}, Engineering Fracture Mechanics
  17~(1) (1983) 1--15.

\bibitem{Murakami1994}
Y.~Murakami, M.~Endo, {Effects of defects, inclusions and inhomogeneities on
  fatigue strength}, International Journal of Fatigue 16~(3) (1993) 163--182.

\bibitem{Crossland1956}
B.~Crossland, {Effect of large hydrostatic pressures on the torsional fatigue
  strength of an alloy steel}, in: The third International Conference on
  Fatigue of Metals, 1956.

\bibitem{Chaboche1989}
J.~L. Chaboche, {Continuum Damage Mechanics: Part I—General Concepts.},
  Journal of Applied Mechanics 55~(1) (1988) 59--64.

\bibitem{Li2023}
P.~Li, W.~Li, B.~Li, S.~Yang, Y.~Shen, Q.~Wang, K.~Zhou, {A review on phase
  field models for fracture and fatigue}, Engineering Fracture Mechanics 289
  (2023) 109419.

\bibitem{Leonard2013}
B.~D. Leonard, F.~Sadeghi, S.~Shinde, M.~Mittelbach, {Rough surface and damage
  mechanics wear modeling using the combined finite-discrete element method},
  Wear 305~(1-2) (2013) 312--321.

\bibitem{Agham2014}
A.~B. Aghdam, A.~Beheshti, M.~M. Khonsari, {Prediction of crack nucleation in
  rough line-contact fretting via continuum damage mechanics approach},
  Tribology Letters 53 (2014) 631--643.

\bibitem{Lorentz2021}
S.~J. Lorenz, F.~Sadeghi, H.~K. Trivedi, L.~Rosado, M.~S. Kirsch, C.~Wang, {A
  continuum damage mechanics finite element model for investigating effects of
  surface roughness on rolling contact fatigue}, International Journal of
  Fatigue 143 (2021) 105986.

\bibitem{Peerlings2000}
R.~H. Peerlings, W.~M. Brekelmans, R.~De~Borst, M.~G. Geers, {Gradient-enhanced
  damage modelling of high-cycle fatigue}, International Journal for Numerical
  Methods in Engineering 49~(12) (2000) 1547--1569.

\bibitem{Marigo2016}
J.~J. Marigo, C.~Maurini, K.~Pham, {An overview of the modelling of fracture by
  gradient damage models}, Meccanica 51 (2016) 3107--3128.

\bibitem{Bourdin2007}
B.~Bourdin, {Numerical implementation of the variational formulation for
  quasi-static brittle fracture}, Interfaces and Free Boundaries 9~(3) (2007)
  411--430.

\bibitem{Alessi2018}
R.~Alessi, S.~Vidoli, L.~De~Lorenzis, {A phenomenological approach to fatigue
  with a variational phase-field model: The one-dimensional case}, Engineering
  Fracture Mechanics 190 (2018) 53--73.

\bibitem{Carrara2020}
P.~Carrara, M.~Ambati, R.~Alessi, L.~De~Lorenzis, {A framework to model the
  fatigue behavior of brittle materials based on a variational phase-field
  approach}, Computer Methods in Applied Mechanics and Engineering 361 (2020)
  112731.

\bibitem{Golahmar2023}
A.~Golahmar, C.~F. Niordson, E.~Mart{\'{i}}nez-Pa{\~{n}}eda, {A phase field
  model for high-cycle fatigue: Total-life analysis}, International Journal of
  Fatigue 170 (2023) 107558.

\bibitem{Cui2024}
C.~Cui, P.~Bortot, M.~Ortolani, E.~Mart{\'{i}}nez-Pa{\~{n}}eda, {Computational
  predictions of hydrogen-assisted fatigue crack growth}, International Journal
  of Hydrogen Energy 72 (2024) 315--325.

\bibitem{Paggi2020}
M.~Paggi, J.~Reinoso, {A variational approach with embedded roughness for
  adhesive contact problems}, Mechanics of Advanced Materials and Structures
  27~(20) (2020) 1731--1747.

\bibitem{Loth2023}
F.~Loth, T.~Kiel, K.~Busch, P.~T. Kristensen, {Surface roughness in
  finite-element meshes: application to plasmonic nanostructures}, Journal of
  the Optical Society of America B 40~(3) (2023) B1--B7.

\bibitem{Novovic2004}
D.~Novovic, R.~C. Dewes, D.~K. Aspinwall, W.~Voice, P.~Bowen, {The effect of
  machined topography and integrity on fatigue life}, International Journal of
  Machine Tools and Manufacture 44~(2-3) (2004) 125--134.

\bibitem{Bourdin2000}
B.~Bourdin, G.~A. Francfort, J.-J. Marigo, {Numerical experiments in revisited
  brittle fracture}, Journal of the Mechanics and Physics of Solids 48~(4)
  (2000) 797--826.

\bibitem{Miehe2010a}
C.~Miehe, F.~Welschinger, M.~Hofacker, {Thermodynamically consistent
  phase‐field models of fracture: Variational principles and multi‐field FE
  implementations}, International Journal for Numerical Methods in Engineering
  83~(10) (2010) 1273--1311.

\bibitem{Freddi2010}
F.~Freddi, G.~Royer-Carfagni, {Regularized variational theories of fracture: A
  unified approach}, Journal of the Mechanics and Physics of Solids 58~(8)
  (2010) 1154--1174.

\bibitem{Lo2019}
Y.~S. Lo, M.~J. Borden, K.~Ravi-Chandar, C.~M. Landis, {A phase-field model for
  fatigue crack growth}, Journal of the Mechanics and Physics of Solids 132
  (2019) 103684.

\bibitem{Pham2011}
K.~Pham, H.~Amor, J.~J. Marigo, C.~Maurini, {Gradient damage models and their
  use to approximate brittle fracture}, in: International Journal of Damage
  Mechanics, Vol.~20, 2011, pp. 618--652.

\bibitem{Mandal2024}
T.~K. Mandal, J.~Parker, M.~Gagliano, E.~Mart{\'{i}}nez-Pa{\~{n}}eda,
  {Computational predictions of weld structural integrity in hydrogen transport
  pipelines}, International Journal of Hydrogen Energy (2024).

\bibitem{Ambati2015}
M.~Ambati, T.~Gerasimov, L.~De~Lorenzis, {A review on phase-field models of
  brittle fracture and a new fast hybrid formulation}, Computational Mechanics
  55 (2015) 383--405.

\bibitem{Gadelmawla2002}
E.~S. Gadelmawla, M.~M. Koura, T.~M.~A. Maksoud, I.~M. Elewa, H.~H. Soliman,
  {Roughness parameters}, Journal of materials processing Technology 123~(1)
  133--145.

\bibitem{Gathimba2019}
N.~Gathimba, Y.~Kitane, T.~Yoshida, Y.~Itoh, {Surface roughness characteristics
  of corroded steel pipe piles exposed to marine environment}, Construction and
  Building Materials 203 (2019) 267--281.

\bibitem{Maradudin1989}
A.~A. Maradudin, T.~Michel, {The transverse correlation length for randomly
  rough surfaces}, Journal of Statistical Physics 58 (1990) 485--501.

\bibitem{Muralikrishnan2009}
B.~Muralikrishnan, J.~Raja, {Computational Surface and Roundness Metrology},
  Springer Science {\&} Business Media, London, 2008.

\bibitem{Blateyron2013}
F.~Blateyron, {The areal field parameters}, in: Characterisation of Areal
  Surface Texture, Springer International Publishing, 2013, pp. 15--43.

\bibitem{ISO2021}
E.~ISO, {Geometrical product specifications (GPS)-Surface texture: Profile-Part
  3: Specification Operators} (2022).

\bibitem{Raoufi2015}
D.~Raoufi, L.~Eftekhari, {Crystallography and morphology dependence of In2O3:
  Sn thin films on deposition rate}, Surface and Coatings Technology 274 (2015)
  44--50.

\bibitem{Montgomery2011}
D.~C. Montgomery, G.~C. Runger, {Applied statistics and probability for
  engineers}, 5th Edition, John Wiley {\&} Sons, 2011.

\bibitem{fenicsx}
M.~Alnaes, J.~Blechta, J.~Hake, A.~Johansson, B.~Kehlet, A.~Logg,
  C.~Richardson, J.~Ring, M.~E. Rognes, G.~N. Wells, {The FEniCS project
  version 1.5}, Archive of numerical software 3~(100) (2015).

\bibitem{gmsh}
C.~Geuzaine, J.~Remacle, {Gmsh: A 3‐D finite element mesh generator with
  built‐in pre‐ and post‐processing facilities}, International Journal
  for Numerical Methods in Engineering 79~(11) (2009) 1309--1331.

\bibitem{Lai1975}
G.~Y. Lai, {On high fracture toughness of coarse-grained AISI 4130 steel},
  Materials Science and Engineering 19~(1) (1975) 153--156.

\bibitem{Dieter1976}
G.~E. Dieter, {Mechanical Metallurgy}, McGraw-Hill, 1976.

\bibitem{Bigerelle2003}
M.~Bigerelle, D.~Najjar, A.~Iost, {Relevance of roughness parameters for
  describing and modelling machined surfaces}, Journal of Materials Science 38
  (1000) 2525--2536.

\bibitem{Roy2018}
S.~Roy, D.~White, S.~Sundararajan, {Correlation between evolution of surface
  roughness parameters and micropitting of carburized steel under boundary
  lubrication condition}, Surface and Coatings Technology 350 (2018) 445--452.

\bibitem{ASTM1}
{ASTM E 466-96}, {Standard practice for Conducting Force Controlled Constant
  Amplitude Axial Fatigue Tests of Metallic Materials}, in: Annual Book of ASTM
  Standards, ASTM International, West Conshohocken, PA, 2021, pp. 466--96.

\bibitem{ASTM2}
{ASTM E 2207-02}, {Standard practice for Strain-Controlled Axial-Torsional
  Fatigue Testing with Thin-Walled Tubular Specimens}, in: Annual Book of ASTM
  Standards, ASTM International, West Conshohocken, PA, 2021, pp. 2207--02.

\bibitem{Kristensen2021}
P.~K. Kristensen, C.~F. Niordson, E.~Mart{\'{i}}nez-Pa{\~{n}}eda, {An
  assessment of phase field fracture: crack initiation and growth},
  Philosophical Transactions of the Royal Society A: Mathematical, Physical and
  Engineering Sciences 379~(2203) (2021) 20210021.

\bibitem{Shrestha2019}
R.~Shrestha, J.~Simsiriwong, N.~Shamsaei, {Fatigue behavior of additive
  manufactured 316L stainless steel parts: Effects of layer orientation and
  surface roughness}, Additive Manufacturing 28 (2019) 23--38.

\bibitem{Thompson2018}
A.~Thompson, N.~Senin, I.~Maskery, L.~K{\"{o}}rner, S.~Lawes, R.~Leach,
  {Internal surface measurement of metal powder bed fusion parts}, Additive
  Manufacturing 20 (2018) 126--133.

\bibitem{Snyder2020}
J.~C. Snyder, K.~A. Thole, {Understanding laser powder bed fusion surface
  roughness}, Journal of Manufacturing Science and Engineering, Transactions of
  the ASME 142~(7) (2020) 071003.

\bibitem{PerezMora2015}
R.~P{\'{e}}rez-Mora, T.~Palin-Luc, C.~Bathias, P.~C. Paris, {Very high cycle
  fatigue of a high strength steel under sea water corrosion: A strong
  corrosion and mechanical damage coupling}, International Journal of Fatigue
  74 (2015) 156--165.

\bibitem{Chauhan2023}
S.~Chauhan, S.~Muthulingam, {Surface roughness characteristics of
  high–ductile thermo–mechanically treated steel rebar exposed to pitting
  corrosion and elevated temperature}, Construction and Building Materials 404
  (2023) 133210.

\end{thebibliography}
\end{document}